\documentclass[letter]{article}

\usepackage[total={6in,8in}]{geometry}
\usepackage{hyperref}
\usepackage[table]{xcolor}

\usepackage{amsmath}
\usepackage{amssymb}
\usepackage{url}
\usepackage{epsfig}
\usepackage{graphicx}
\usepackage{algorithm}
\usepackage{tikz}

\usepackage[noend]{algorithmic}
\usepackage{rotating}
\usepackage{subfig}
\usepackage{color}
\usepackage{multirow}
\usepackage{amsmath}
\usepackage{xspace}
\usepackage{bm}

\usepackage{enumitem}
\usepackage{lipsum}
\usepackage{setspace}
\usepackage[font=footnotesize,labelfont=bf]{caption}
\let\OLDthebibliography\thebibliography
\renewcommand\thebibliography[1]{
  \OLDthebibliography{#1}
  \setlength{\parskip}{0pt}
  \setlength{\itemsep}{0pt plus 0.3ex}
}

\newcommand\verysmallfont{\fontsize{10}{11}\selectfont}
\newcommand\verysmallerfont{\fontsize{10}{11}\selectfont}
\newcommand\extremelysmallfont{\fontsize{10}{11}\selectfont}

\newcommand{\myspace}[1]{}
\newcommand\spgemm{{\sc spgemm}}
\newcommand\mkl{{\sc mkl}}
\newcommand\gpu{{\sc gpu}}
\newcommand\knl{{\sc knl}}
\newcommand\cpu{{\sc cpu}}

\newcommand\flops{$f_m$}
\def\cca#1{\cellcolor{black!#10}\ifnum #1>5\color{white}\fi{#1}}
\newcommand{\Lone}{$\mathcal{L}_1$}
\newcommand{\Ltwo}{$\mathcal{L}_2$}
\newcommand\maxrowsize{{\sc maxrs}}
\newcommand\avgrowsize{{\sc ars}}
\newcommand\maxrowflops{{\sc maxrf}}

\newcommand{\mdcomment}[1]{{\color{black}{#1}}}

\begin{document}

%\myspace{-3ex}
\title{Multi-threaded Sparse Matrix-Matrix Multiplication for Many-Core and GPU Architectures
}

\author{Mehmet Deveci, Christian Trott, Sivasankaran Rajamanickam\\ \{mndevec,crtrott,srajama\}@sandia.gov \\ Sandia National Laboratories, Albuquerque, NM \\ SAND2018-0186 R}
%\myspace{-25ex}

\maketitle

%\cortext[mycorrespondingauthor]{Corresponding author. \{mndevec,crtrott,srajama\}@sandia.gov}
%\ead{\{mndevec,crtrott,srajama\}@sandia.gov }

\onehalfspacing
%\myspace{-3ex}
\begin{abstract}
%\myspace{-3ex}
%\begin{abstract} \small\baselineskip=9pt 

Sparse Matrix-Matrix multiplication is a key kernel that has applications in
several domains such as scientific computing and graph analysis. 
Several algorithms have been studied in the past for this foundational kernel.
In this paper, we develop parallel algorithms for 
sparse matrix-matrix multiplication 
with a focus on performance portability across different high performance computing architectures.
%with a focus on three different 
%architectures, IBM Power 8, Intel Xeon Phi, and NVIDIA GPUs. 
The performance
of these algorithms depend on the data structures used in them. We compare
different types of accumulators in these algorithms and
demonstrate the performance difference between these data structures.
Furthermore, we develop a meta-algorithm, {\sc kkSpGEMM}, to choose the
right algorithm and data structure based on the characteristics of the 
problem. We show performance comparisons on
three architectures and demonstrate the need for the community to 
develop two phase sparse matrix-matrix multiplication implementations
for efficient reuse of the data structures involved.

\end{abstract}
\myspace{-3ex}
%\begin{keyword}
%Sparse Matrix-Sparse Matrix Multiplication, GPU, KNL, multi-threading
%\end{keyword}

%\linenumbers

%\maketitle

%\thanks{
%This work is supported by the U.S. Dept. of Energy, 
%Office of Science, Office of Advanced Scientific Computing 
%Research, Scientific Discovery through Advanced Computing 
%(SciDAC) program, and by the NNSA’s Advanced Simulation 
%and Computing (ASC) program.
%}

\date{}

%\maketitle

%\pagenumbering{arabic}
%\setcounter{page}{1}%Leave this line commented out.

\myspace{-5ex}

\myspace{-0.5ex}

\section{Introduction}
\label{sec:intro}
\myspace{-3ex}

Modern supercomputer architectures are following various different
paths, e.g., Intel's XeonPhi processors, NVIDIA's Graphic Processing Units ({\gpu{}s}) or 
the Emu systems~\cite{dysart2016highly}.
Such an environment increases the importance of designing 
flexible algorithms for performance-critical kernels and implementations 
that can run well on various platforms. 
We develop multi-threaded algorithms for sparse matrix-matrix multiply 
(\spgemm{}) kernels in this work.
\spgemm{} is a fundamental kernel that is used in various applications such as graph
analytics~\cite{wolf2017fast} and scientific computing, especially in the setup
phase of multigrid solvers~\cite{lin2014towards}. The kernel has been studied
extensively in the contexts of sequential~\cite{gustavson1978two}, shared
memory parallel~\cite{patwary2015parallel, intel2007intel} and
\gpu{}~\cite{demouth2012sparse, liu2014efficient, gremse2015gpu,
dalton2015optimizing} implementations.  There are optimized kernels available on
different architectures~\cite{intel2007intel, liu2014efficient,
dalton2015optimizing, Rupp:ViennaCL, naumov2010cusparse} providing us with good
comparison points.
%The focus on this problem has been around programming models to implement 
%an algorithm on multiple architectures~\cite{edwards2014kokkos,openmpgpus}.
%We consider this problem from an algorithmic perspective to design a 
%``performance-portable algorithm'', an algorithm that can perform well 
%on multiple architectures with similar accuracy and robustness with
%minimal assumptions about the underlying architecture.
%Another approach would be to consider different, architecture specific 
%(``native'') algorithms that is highly optimized with the specifics of the
%architecture for every key kernel. Although this might help to achieve higher
%performance for a given architecture, it requires revisiting algorithmic design
%with significant amount of changes for each different architecture. 
\mdcomment {In this work, we provide portable algorithms for the \spgemm{} kernel 
and their implementations using Kokkos~\cite{edwards2014kokkos} programming model with minimal changes for 
the architectures' very different characteristics. 
For example, traditional \cpu{}s have powerful cores 
with large caches, while XeonPhi processors have many lightweight cores, 
and GPUs provide extensive hierarchical parallelism with very simple computational units.}
%The algorithmic changes required are focused on the assignment of computation to
%execution units (partitioning scheme) and data structures used in different
%architectures.
%MDNOTE:NOTE SURE ABOUT THE ABOVE SENTENCE.
The algorithms in this paper aim to minimize revisiting algorithmic design 
for these different architectures. 
%Using the Kokkos programming model, the code divergence in the implementation
%are limited to access strategies of different data structures and how different levels 
%of parallelism in the algorithm are mapped to computational units.
%MD: Karen's update:
The code divergence in the implementation %handles different data structures 
and how different levels of algorithmic parallelism are mapped to computational units.
%and computation unit hierarchy in different architectures.
is limited to access strategies of different data structures and how different levels 
of parallelism in the algorithm are mapped to computational units.

%This leads to the question ``How much performance will be sacrificed for portability?''. We address this question
%by comparing the implementation of our performance-portable algorithm to 
%several native implementations.

%It is important to study this problem with a kernel which is reasonably
%difficult and uses standard patterns, so one can reason about designing
%other kernels based on the knowledge gained. 

An earlier version of this paper~\cite{deveci2017performance} focuses on \spgemm{} from the perspective of
performance-portability. 
It addressed the issue of performance-portability for \spgemm{} with 
an algorithm called {\sc kkmem}.
%and a two-level sparse hashmap accumulator with a memory-pool.
It demonstrated better performance on \gpu{}s 
and the current generation of XeonPhi processors, Knights Landing (\knl{}s),
w.r.t. state-of-art libraries. 
%We have extended the \spgemm{} implementation with algorithms that use other
%partitioning schemes and data structures. 
%\begin{itemize} [topsep=0pt,itemsep=-1ex,parsep=1ex,leftmargin=1pt,	itemindent=10pt]
%\item What are the performance critical design choices
%and data structures for the \spgemm{} algorithm to map well to 
%different architectures (thousands vs hundreds of threads, 
%streaming multiprocessors vs lightweight cores, 
%shared memory vs {\sc mcdram}) ? 
%%(HBM or {\sc mcdram}))? 
%%	\item What are the data structures needed to realize an efficient
%%	implementation of this algorithm ? 
%
%\item How will the kernel serve the needs of real applications, when
%there is a reuse of the symbolic structure?%, or when the architecture 
%%has limited HBM?
%
%%\item Can we develop a theoretical model for memory accesses of the algorithms 
%%to understand the
%%performance of the kernel with different memory types?
%
%\end{itemize}
Our contributions in~\cite{deveci2017performance} is summarized below.
\begin{itemize} %[leftmargin=*,topsep=0pt,itemsep=-1ex,partopsep=1ex,parsep=1ex,leftmargin=0pt,	itemindent=10pt]

\item We design two thread-scalable data structures (multilevel hashmap accumulators 
and a memory pool) to achieve scalability on various platforms, and a graph compression 
technique to speedup the symbolic factorization of \spgemm{}. 

\item We design hierarchical, thread-scalable \spgemm{} algorithms and implement them 
using the Kokkos programming model.
% and compare its performance to other native implementations. 
Our implementation is  available at \\
\texttt{https://github.com/kokkos/kokkos-kernels} and also in the 
Trilinos framework \\ (\texttt{https://github.com/trilinos/Trilinos}).

\item We also present results for the practical case of matrix structure reuse, 
and demonstrate its importance for application performance.
\end{itemize}

\noindent
%This extended version of the conference paper considers several new algorithm design 
%choices, additional data structures, and their practical usage in a two-level \spgemm{} method.
This paper extends~\cite{deveci2017performance} with several new algorithm design choices and
additional data structures. Its contributions are summarized below.

\begin{itemize} %[leftmargin=*,topsep=0pt,itemsep=-1ex,partopsep=1ex,parsep=1ex,leftmargin=0pt,	itemindent=10pt]

\item We present results for the selection of kernel parameters e.g., partitioning scheme 
and data structures with trade-offs for memory access vs. computational overhead cost, 
and provide heuristics to choose the best parameters depending on the problem 
characteristics.

\item We extend the evaluation of the performance of our methods on various platforms, 
including traditional \cpu{}s, \knl{}s, and \gpu{}s.
We show that our method achieves better performance than native methods on IBM Power8 \cpu{}s, 
and \knl{}s. It outperforms two other native methods on \gpu{}s, and achieves 
similar performance to a third highly-optimized implementation.

\end{itemize}

%\item We develop a theoretical hypergraph model for measuring memory accesses
%in both architectures and study its relationship to measured performance. 

%\todo{It would be nice to separate the above list to this paper's contributions. It would be good to properly indent the items too.}
%MD: I will work on indentation at the end. Curentlystill need to cut 3 pages.

The rest of the paper is organized as follows: Section~\ref{sec:background}
covers the background for \spgemm{}. 
%especially focusing on important differences between the different native implementations. 
Our \spgemm{} algorithm and
related data structures are described in Section~\ref{sec:algo}.
%The theoretical model for memory accesses is given in
%Section~\ref{sec:hypergraph}. 
Finally, the performance comparisons that
demonstrate the efficiency of our approach is given in Section~\ref{sec:results}.

\myspace{-4.5ex}
\section{Background}
\label{sec:background}
\myspace{-2.5ex}

%In this work, we study scalable and memory efficient \spgemm{} methods for
%multicore and many-core architectures. 
Given matrices $A$ of size $m \times n$ and $B$ of size 
$n \times k$ \spgemm{} finds the
$m \times k$ matrix $C$ s. t. $C = A \times B$.
Multigrid solvers use triple
products in their setup phase, which are of the form $A_{coarse} = R \times
A_{fine} \times P$ ($R = P^T$ if $A_{fine}$ is symmetric), to coarsen the
matrices. \spgemm{} is also widely used for various graph analytic problems~\cite{wolf2017fast}. 

\begin{algorithm}
\caption{
\verysmallfont
%Row-wise 
\spgemm{} for $C=A \times B$. $C(i,:)$ 
($C(:,i)$) refer to $i^{th}$ row (column) of $C$.}

\begin{algorithmic}[1]
\verysmallfont

\REQUIRE{Matrices $A$, $B$}
\FOR {\textcolor{red} {$i \gets 0$ to $m - 1$}}  \label{ln:gust1d} 
  \FOR {\textcolor{blue} {$j \in A(i,:)$}}	\label{ln:gust2d} 
    \STATE {\tt //accumulate partial row results}
    \STATE {\textcolor{green} {$C(i,:) \gets C(i,:) + A(i, j) \times B(j,:)$}} \label{ln:gustacc}
  \ENDFOR
\ENDFOR
\end{algorithmic}
\label{alg:spgemm1d}
\end{algorithm}

%In the literature, 
Most parallel \spgemm{} methods follow Gustavson's
algorithm~\cite{gustavson1978two} (Algorithm~\ref{alg:spgemm1d}).  This
algorithm iterates over rows of $A$ %in a 1D fashion
(line~\ref{ln:gust1d}) to compute all entries in the corresponding row of $C$. 
Each iteration of the second loop (line~\ref{ln:gust2d})
{\emph {accumulates the intermediate values}} of multiple columns within the row using an accumulator.
%The number of the necessary multiplications to perform this matrix multiplication 
The number of multiplications needed to perform this matrix multiplication 
is called \flops{} %{\sc flops} 
(there are \flops{}  additions too) for the rest of the paper. 
%\todo{Need to use a different name than {\sc flops} here. MFLOPS will be confusing as well.
%M\_FLOPS or $\mbox{flops}_m$? or $f_m$. Let us change FLOPS is per second and GFLOPS/per second
%everywhere. That is the more common thing}.
%MD: Added a formula for this.

%\lipsum[3-6]
%\setlength{\textfloatsep}{0pt}% Remove \textfloatsep

\noindent
%\paragraph
\textbf{Design Choices:} There are three design choices that can be made in 
Algorithm~\ref{alg:spgemm1d}:
(a) the partitioning needed for the iterations, (b) how to determine the size of $C$ as it 
is not known ahead of time, and (c) the different
\mdcomment {data structures} for the accumulators. 
The key differences in past work are related to
these three choices in addition to the parallel programming model.
%(distributed memory vs shared memory).
%This section gives a brief background of the different choices and what is
%explored in the literature. % with a summary of different methods listed
%in Table~\ref{tab:literature}.

%Different partitioning schemes have been used in the past for {\sc spgemm}. 
\mdcomment{First design choice is how to distribute the computation over execution units.} % in the literature.}
A 1D partitioning method~\cite{heroux2005overview} partitions $C$ along a single dimension, and 
each row is computed by a single execution unit. On the other hand, 
2D~\cite{patwary2015parallel,bulucc2011combinatorial} and 
3D~\cite{azad2015exploiting} methods assign each nonzero of $C$ 
or each multiplication to a single execution unit, respectively. 
Hypergraph partitioning methods have also been used to improve
the data locality in 1D~\cite{akbudak2014simultaneous,akbudak2017exploiting} 
and 3D~\cite{ballard2016hypergraph} methods.
1D row-wise is the most popular choice for scientific computing applications.
Using partitioning schemes for \spgemm{} that differ from the application's scheme requires 
reordering and maintaining a copy of one or both of the input matrices.
%, if the same scheme is not used throughout the rest of the application.
% Large number of matrix multiplies are needed
%to amortize the reordering costs.
For \gpu{}s, hierarchical algorithms are also employed, where rows are assigned to a first level 
of parallelism (blocks or warps), and the calculations within the rows are done using 
a second level of parallelism~\cite{demouth2012sparse, Rupp:ViennaCL, liu2014efficient,dalton2015optimizing}. 
In this work, \emph{we use such a hierarchical partitioning of the computation, where the 
first level will do 1D partitioning and the second level will exploit further 
thread/vector parallelism.}
%At the vector level the algorithm behaves like a 2D algorithm

The second design choice is how to determine the size of $C$. Finding the structure of $C$ is usually 
as expensive as finding $C$. 
There exists some work to
estimate its structure~\cite{cohen1998structure}. However, it does not provide
a robust upper bound and it is not 
%It also requires a generation of a random dense matrix, and 
%a sparse matrix-dense matrix multiplication ({\sc spmm}) which is not 
significantly cheaper than calculating the exact size in practice.
As a result, both one-phase and two-phase methods are commonly used. 
One-phase methods rely either on finding an upper bound for the size 
of $C$~\cite{kurt2017Characterization} or doing dynamic reallocations when needed. 
The former could result in over-allocation and the
latter is not feasible for \gpu{}s. 
Two-phase methods first compute the structure of $C$  (\emph {symbolic} phase), 
before computing its values in the second phase (\emph {numeric} phase). 
%Two-phase methods use the structure 
%of $A$ and $B$ to determine or estimate the structure of $C$  (\emph {symbolic} phase), 
%before computing $C$ in the second phase (\emph {numeric} phase). 
%They either find the size of the rows of $C$ or its
%exact non-zero pattern~\cite{naumov2010cusparse, demouth2012sparse}, and 
%They find the size of the rows of $C$, and 
They allow reusing the structure $C$ for different multiplies with the same structure 
of $A$ and $B$~\cite{naumov2010cusparse, demouth2012sparse}. This is an important use case 
in scientific computing, where matrix structures stay constant while matrix values change 
frequently~\cite{deveci2017performance}. 
%In such problems, the symbolic phase %of a two-phase 
%\spgemm{} method 
%can be executed once, with its result reused multiple times for matrices 
%with changing values. 
The two-phase method also provides significant advantages in graph analytics. 
Most of them work only on the symbolic structure, skipping the numeric phase~\cite{wolf2017fast}. %For example, our recent work on the triangle counting problem~\cite{wolf2017fast} 
%uses only \spgemm{}'s symbolic phase. 
In this work, \emph{we use a two-phase approach, and 
speed the symbolic phase up using matrix compression}.

The third design choice is the data structure to use for the accumulators. 
Some algorithms use a dense data structure of size $k$. %the number of columns in $B$ ($k$). 
The intermediate results for a row are stored in an array of size $k$ in its ``dense'' format. 
%While such methods do not incur any overhead for index calculations,
These dense thread-private arrays may not be scalable for massive amounts of threads 
and large $k$ values.
%Moreover, random accesses
%to such large arrays might harm the overall performance because of the memory
%bandwidth issues. 
Therefore, sparse accumulators such as heaps or hashmaps are
preferred. In this work, 
\emph{we use both multi-level hashmaps as sparse accumulators and dense accumulators 
to achieve scalability in \spgemm{}.}

%There are various works that consider distributed parallization of \spgemm{}. 
%Tpetra package of Trilinos~\cite{heroux2005overview} performs $1D$ row-wise 
%partitioning that is based on the Gustavson's algorithm. In a $1D$ algorithm
%the result matrix is $1D$ partitioned and each row is calculated by a single 
%computation unit. On the other hand, SUMMA algorithm that partitions the result matrix 
%into $2D$ space is adopted in Combinatorial BLAS~\cite{bulucc2011combinatorial}. This algorithm
%is extended to $3D$, via parallelization over calculation of single entry 
%in~\cite{azad2015exploiting}. Hypergraph models and algorithms that use
%hypergraph partitioning for sparse matrix-matrix outer-product multiplications 
%and $3D$ computation partitioning are studied in~\cite{akbudak2014simultaneous} 
%and in~\cite{ballard2016hypergraph}, respectively.

%\paragraph
\noindent
\textbf{Related Work:} There are a number of distributed-memory algorithms for 
\spgemm{}~\cite{bulucc2011combinatorial,akbudak2014simultaneous,ballard2016hypergraph,azad2015exploiting,heroux2005overview}.
Most of the multithreaded \spgemm{} studies ~\cite{patwary2015parallel,
azad2015exploiting,Rupp:ViennaCL,gremse2015gpu,intel2007intel}
%are based on Gustavson's algorithm.
follow Gustavson's algorithm, and  
%They study efficient parallelization of Gustavson's method, and 
%They usually 
differ in the data structure used for row accumulation. 
Some use dense accumulators~\cite{patwary2015parallel},
others a heap with an assumption of sorted columns in $B$ rows~\cite{azad2015exploiting},
or sorted row merges~\cite{Rupp:ViennaCL,gremse2015gpu}.

%One option is to use dense
%accumulators~\cite{patwary2015parallel}, which have the size of columns of the
%result matrix. 
%Another approach is the use of sparse accumulators. Azad et.
%al.~\cite{azad2015exploiting} uses heaps as accumulators in their shared memory
%implemention.  This bases on the assumption that the rows of $B$ have sorted
%column indices. 
%Similarly, ViennaCL~\cite{Rupp:ViennaCL} provides SPGEMM 
%implementations written with CUDA, OpenCL and OpenMP, and it implements row
%merges~\cite{gremse2015gpu} that bases on merge sort of the columns of $B$.
%\mkl{}~\cite{intel2007intel} also provides a multi-threaded implementation that
%uses sparse accumulators without any assumption on the order of the columns of
%$B$.  
%As a different approach than all, Patwary et.
%al.~\cite{patwary2015parallel} also studies $2D$ partitioning of the result
%matrix. At their preprocessing step, they partition $B$ based on the columns,
%however the gain only amortizes the cost of preprocessing on certain
%conditions, for which they have a heuristic for partitioning decision.

%Extensive parallelism of \gpu{}s are exploited for \spgemm{} in the literature, and most of the algorithms
Most of the \spgemm{} algorithms for \gpu{}s are hierarchical. 
%and use multiple levels of partitioning. 
CUSP~\cite{dalton2015optimizing} uses a hierarchical algorithm 
%that is based on expand and sort methods (ESC). 
where each multiplication is computed
by a single thread and later accumulated with a sort operation. % (ESC). 
%Such global expansion results in high memory requirements. 
AmgX~\cite{demouth2012sparse} follows a hierarchical 
Gustavson algorithm.  Each row is calculated by a single warp, and multiplications 
within a row are done by different threads of the warp. It uses $2$-level cuckoo-hash accumulators, 
and does not make any assumption on the order of the column indices.
On the other hand, the row merge algorithm~\cite{gremse2015gpu} and its implementation in
ViennaCL~\cite{Rupp:ViennaCL} uses merge sorts for accumulations of the sorted rows. 
bhSPARSE~\cite{liu2014efficient} also exploits this assumption on \gpu{}s. 
%It has methods
%to predict the size of the result matrix. Exploiting that knowledge, it performs binning based on the size of the 
%result rows and chooses different accumulators based on the size of the row.
It chooses different accumulators based on the size of the row.
A recent work Nsparse~\cite{nagasaka2017high} also employs a hierarchical 
method and uses linear probing for accumulations. It places rows into bins
based on the required number of multiplications and the output row size, 
and launches different concurrent kernels for each bin. 
%It is one of the 
%fastest methods to our knowledge in the current literature.
%and performs accumulation for the multiplication of the row 
%using heap accumulators, ESC or row merge based on the size of the row. 
Different from most of the \spgemm{} work, McCourt et. al~\cite{mccourt2013efficient} 
computes a distance-2 graph coloring on the structure of $C$ in order to reduce \spgemm{}
to {\sc spmm}.
\noindent
\textbf {Kokkos:}
Kokkos~\cite{edwards2014kokkos} is a C++ library providing an abstract
data and task parallel programming model, which enables performance portability
%for high performance computing architectures. It provides a
for various architectures. It provides a
single programming interface but allows different optimizations for backends
such as OpenMP and {\sc cuda}.
%We use Kokkos' features such as
%parallel\_for, parallel\_scan, atomics, and views (arrays), and 
%the Kokkos thread hierarchy for hierarchical parallelism. 
Using Kokkos enables us to run the same code on the \cpu{}s, 
\knl{}s and \gpu{}s just compiled differently.

%\begin{table}[]
%\centering
%\caption{\verysmallfont Kokkos Hierarchy mapping to \gpu{}s and \knl{}s/\cpu{}s}
%\label{tab:kokkoshierarchy}
%\resizebox{\columnwidth}{!}{%
%\begin{tabular}{c|c|c}
%KK \& Kokkos & \gpu{}s & \knl{}s/\cpu{}s \\ \hline
%Team & Thread Block & Work assigned to group of hyperthreads \\ \hline
%Kokkos Thread & (full, half, quarter...) Warp & Work assigned to a single Thread \\ \hline
%Vector Lane & Threads within a warp & Vectorization Units \\ 
%\end{tabular}
%}
%\end{table}

The kokkos-parallel hierarchy consists of teams, threads and vector lanes. 
%Table~\ref{tab:kokkoshierarchy} shows how these terms map to execution 
%units on \gpu{}s, \knl{}s and \cpu{}s. The mapping is the same for \knl{}s
%and \cpu{}s. 
A \emph{team} in Kokkos handles a workset assigned 
to a group of threads sharing resources. On \gpu{}s, it is mapped to a thread block,
which has access to a software managed L1 cache.
%A team on the \knl{} 
%can be all threads sharing the {\sc ddr} memory, or the threads sharing an L2 or
A team on \cpu{}s (or \knl{}s) is a collection of threads sharing some common resources.
Depending on the granularity of the work units,
a team is commonly chosen as the group of hyperthreads 
that share an L2/L1 cache or even just a single hyperthread.  
In this work, we use a team size of one (a single hyperthread) 
on \cpu{}s. \mdcomment{On \gpu{}s, a typical team size is between $4$ and $64$. } 
%chosen based on the number of warps in a thread block.}
%We use the hyperthreads that share an L1 cache as the team.
%In our experience, the best performance with teams is achieved by 
%using the hyperthreads that share an L1 cache. 
%Work-sharing of the threads within the 
%team is achieved by assigning consecutive indices to consecutive threads. 
%This allows coalesced memory accesses on \gpu{}s and better cache reuse on \cpu{}s
%if teams are chosen as hyperthreads. 
%Note that, if a team on \cpu{} consists of threads that shares {\sc ddr} memory, this causes false
%sharing. 
There is no one-to-one mapping from teams to 
the number of execution units. That is, the number of teams, even on \cpu{}s, can be much higher than the number of 
execution units. 
It is therefore useful to think of teams as a logical concept, with a one-to-one mapping to work items. 
A \emph{kokkos-thread} within a team  maps to 
a warp or warp fraction (half, quarter, etc.) on \gpu{}s and to a single thread on
\cpu{}s. A kokkos-thread uses multiple vector lanes, which %The \emph{vector lanes}
map to cuda-threads within a warp in \gpu{}s  and the vectorization units on \cpu{}s. 
The length of the vector lanes, {\em vector length}, is a runtime parameter on \gpu{}s 
and can be at most the length of a warp, while on \cpu{}s it is fixed depending on the 
architecture. We use the terms teams, {\em threads (for kokkos-threads)} and 
vector lanes in the rest of the paper. %, which refers to the explained computational units on \cpu{}s and \gpu{}s. 

\noindent
The portability provided by Kokkos comes with some overhead. 
For example, heavily used template meta-programming %used heavily in Kokkos 
causes some compilers to fail to perform certain optimizations. 
%Kokkos wraps 
%backends such as Cuda and OpenMP, and compilers are confused with such new-technology wrappers.
%Typically, portable data structures in Kokkos have some small overheads as well.
Portable data structures have also small overheads.
%Portable methods also avoid exploiting any architecture-specific assumption
%or instructions. However, all these overheads can be overcome by careful
%algorithm design unless the problem sizes are very small. 
While Kokkos allows us to write portable kernels, complex ones as \spgemm{}
can benefit from some code divergence for better performance. 
For example, our implementations favor atomic operations on \gpu{}s, and reductions on 
\cpu{}s. 

\myspace{-3.5ex}
\section{Algorithms}
\label{sec:algo}
\myspace{-1.25ex}

\begin{algorithm}
\caption{\verysmallfont Overall structure of {\sc SpGEMM} Methods.}
\begin{algorithmic}[1]
\verysmallfont
\REQUIRE{Input matrices $A$, $B$ s.t. $C=A\times B$}
\STATE {\tt allocate $C_{\tt row\_pointers}$ }
\STATE {\tt $B_c \gets $ {\tt compress\_matrix}$(B)$ }
\STATE {\tt $C_{\tt row\_pointers} \gets $ {\tt {\sc core\_spgemm}} $(`symbolic',A,B_c)$ {\tt //symbolic phase}}
\STATE {\tt allocate $C_{\tt columns}$, $C_{\tt values}$}
\STATE {\tt $C \gets $ {\tt {\sc core\_spgemm}} $(`numeric',A,B,C_{\tt row\_pointers})$ {\tt //numeric phase}
}

\end{algorithmic}
\label{alg:kkspgemm}
\end{algorithm}

\noindent
%This section explains the overall structure of our \spgemm{} methods
The overall structure of our \spgemm{} methods is given in Algorithm~\ref{alg:kkspgemm}.
It consists of a two-phase approach, in which the first (symbolic) phase computes the number of nonzeros
in each row (line $3$) of $C$, and the second (numeric) phase
(line $5$) computes $C$. Both phases use the {\tt {\sc core\_spgemm}} kernel with small changes.
The main difference of the two phases is that the symbolic phase does not use the matrix values,
and thus performs no floating point operations. We aim to improve memory and runtime of the symbolic phase
by compressing $B$. % (Line $2$).
%this phase and 
%reduce memory usage by compressing the ordinals of $B$ (Line $2$).
%there is no read, multiply and accumulation of the floating points. However, in order 
%to improve this phase and reduce the memory usage of accumulators, we compress 
%$B$ into a smaller data structure (Line 4), which introduces new bit values
%(see Section~\ref{sec:compression}).
%The core \spgemm{} kernel, and the compression method 
%used in {\sc kkmem} are described below.

\myspace{-2ex}
\subsection{Core \spgemm{} Kernel} 
\myspace{-1.25ex}

The core \spgemm{} kernel used by the symbolic and the numeric phase uses a hierarchical, 
row-wise algorithm (\ref{alg:spgemmcoregeneral}) 
with two thread-scalable data structures: a memory pool and an accumulator.
A team of threads, which depending on the architecture may be a single thread or many, 
is assigned a set of rows over which it loops. 
For each row $i$ of $A$ within the assigned rows, we traverse
the nonzeroes $A(i,j)$ of $A(i,:)$ (Line $4$). 
Column/Value pairs of the corresponding row of $B(j,:)$ are multiplied and 
inserted (\mdcomment {either as a new value or accumulated to an existing one}) 
into a small-sized level-1 (\Lone{}) accumulator. 
\Lone{} is located in 
fast memory and allocated using team resources (e.g., shared memory on \gpu{}s). 
If \Lone{} runs out of space, the partial results are inserted into a level-2 
(\Ltwo{}) accumulator located in global memory.

\begin{algorithm}
\caption{\verysmallfont 
{\sc core\_spgemm} Kernel for $C=A \times B$. 
Based on the phase (symbolic/numeric), 
$B$ is either a compressed or standard matrix. Either $C_{\tt row\_pointers}$ or $C$ is filled.
%Colors show how the hierarchical algorithm 
%maps to Algorithm~\ref{alg:spgemm1d}. Red shows the first level, 
%and green shows the second level of parallelism. Blue part is 
%sequential within the first level parallelism.
}
\begin{algorithmic}[1]
\verysmallfont
\REQUIRE{Phase, Matrices $A$, $B$, $C$.}
%\STATE {\tt // based on phase (symbolic/numeric)  }
%\STATE {\tt // $B$ is either a compressed or standard matrix }
%\STATE {\tt // and either $C_{\tt row\_pointers}$ or $C$ is filled.}
  \STATE {allocate the first level accumulator \Lone{}}
  \STATE {$TeamRows \gets$ {\sc getTeamRows}(thread\_team)}
  \FOR {\textcolor{red} {$i$ $\in$ $TeamRows$}}
    \FOR {\textcolor{blue} {j $\in $ $A(i,:)$}}
      \FOR {\textcolor{green} {$col, val$ $\in$ {\sc $B(j,:)$}}}
        \STATE {\textcolor{black} {$tmpval \gets$ {\sc $val \times A(i,j)$)}}}
         \IF {{\sc Full} \textcolor{black} {$ = $\Lone{}.{\sc insert}$(col, tmpval)$}}
           \IF {\Ltwo{} is not allocated}%$L2\_not\_allocated$}
	    \STATE allocate the second level accumulator \Ltwo{}
            %\STATE $L2\_allocated \gets True$
          \ENDIF
          \STATE \textcolor{black} {\Ltwo{}.{\sc insert}$(col, tmpval)$}
         \ENDIF
    \ENDFOR
  \ENDFOR

    %\IF {{\sc Phase is Symbolic}}       \STATE $C_{\tt row\_pointers}(i) \gets $ total \Lone{}/\Ltwo{} Acc sizes
    \STATE {\textbf{if} {\sc Phase is Symbolic} \,\,\,\,\,\,\,\,\, {\textbf{then}}} $C_{\tt row\_pointers}(i) \gets $ total \Lone{}/\Ltwo{} Acc sizes
    \STATE {\textbf{else} {\textbf{if} {\sc Phase is Numeric} {\textbf{then}}} } $C(i,:) \gets $ values from \Lone{}/\Ltwo{} Acc
%    \ELSE
%      \STATE $C(i,:) \gets $ values from \Lone{}/\Ltwo{} Acc
%    \ENDIF
    \STATE {reset \Lone{}, release \Ltwo{} if allocated.}
%    \IF {$L2\_allocated$}
%          \STATE release \Ltwo{}
%    \ENDIF
\ENDFOR

\end{algorithmic}
\label{alg:spgemmcoregeneral}
\end{algorithm}

First, we focus on partitioning the computation using hierarchical parallelism. 
The first level parallelism is trivially achieved by assigning disjoint 
sets of rows of $C$ to teams (Line $2$). Further parallelization can be achieved
on the three loops highlighted with red, blue and green (Lines $3$, $4$ and $5$).
Each of these loops can either be executed sequentially by the whole processing unit 
(team), or be executed in parallel by partitioning over threads of the teams.

\myspace{-3ex}
\subsubsection{\spgemm{} Partitioning Schemes}
\myspace{-1.25ex}

\begin{figure}[t!]
    \centering
     %\subfloat[thread hierarchy]
    %{}
    \includegraphics[width=0.40\columnwidth]{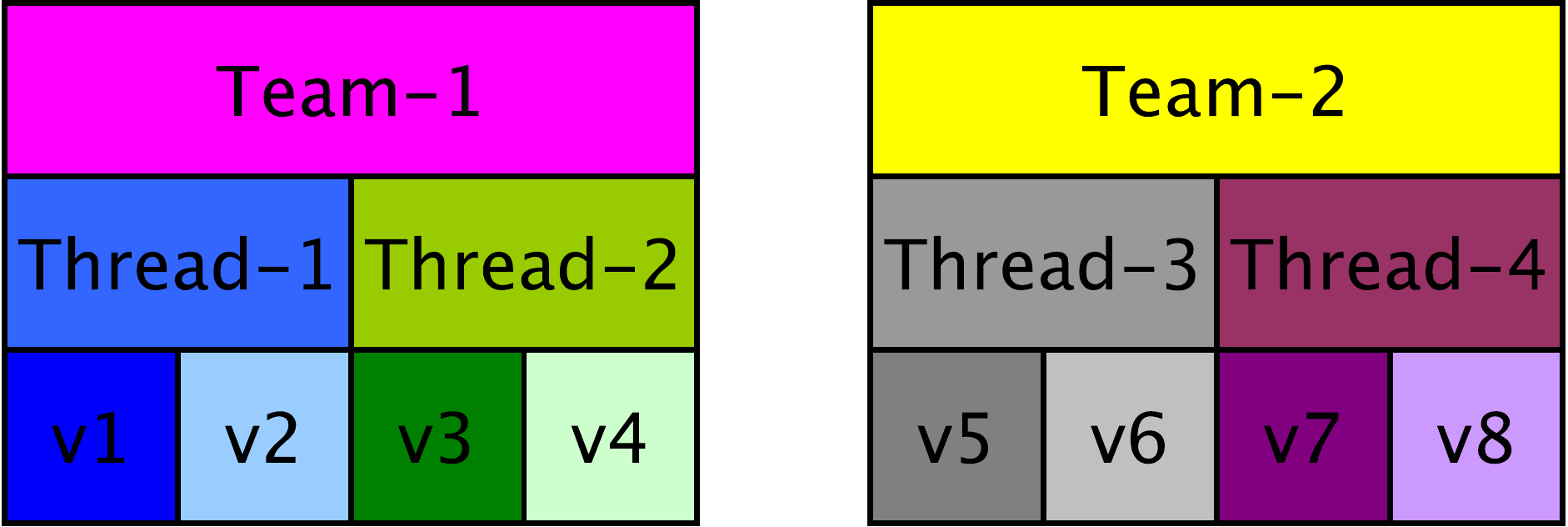} 
    \caption{\verysmallfont Thread hierarchy used in Figure~\ref{fig:spgemmhierarchy} and Figure~\ref{fig:spgemmhierarchy2}. 
    Two teams have two threads with two vector lanes each.}
    \label{fig:hier}
\end{figure}

\begin{figure}[t!]
    \centering
     \subfloat[\extremelysmallfont Thread-Sequential: Thread-1 is assigned to a single row of $A$. 
				  It sequentially traverses the corresponding rows of $B$, one and six.
				  It exploits vector parallelism for rows of $B$.]
    {\includegraphics[width=1\columnwidth]{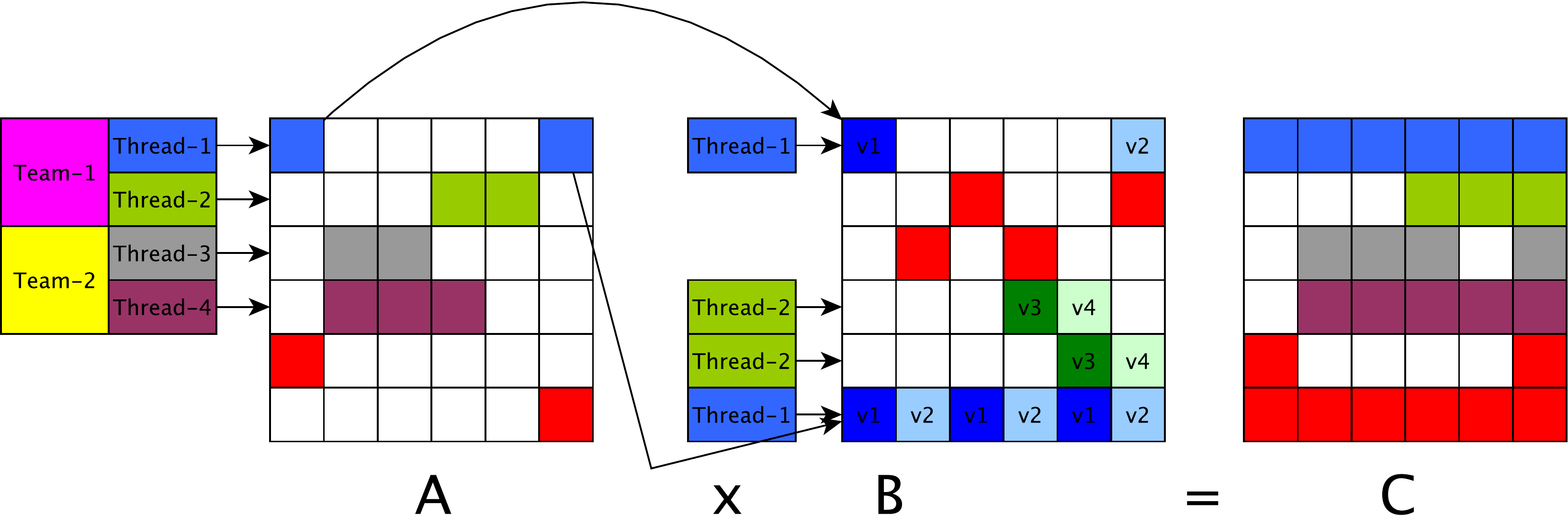}
    \label{fig:threadseq}}

     \subfloat[\extremelysmallfont Team-Sequential:Team-1 is assigned to a single row of $A$. 
			       It sequentially traverses the corresponding rows of $B$,  one and six.
			       It exploits both thread and vector parallelism for rows of $B$.]
    {\includegraphics[width=1\columnwidth]{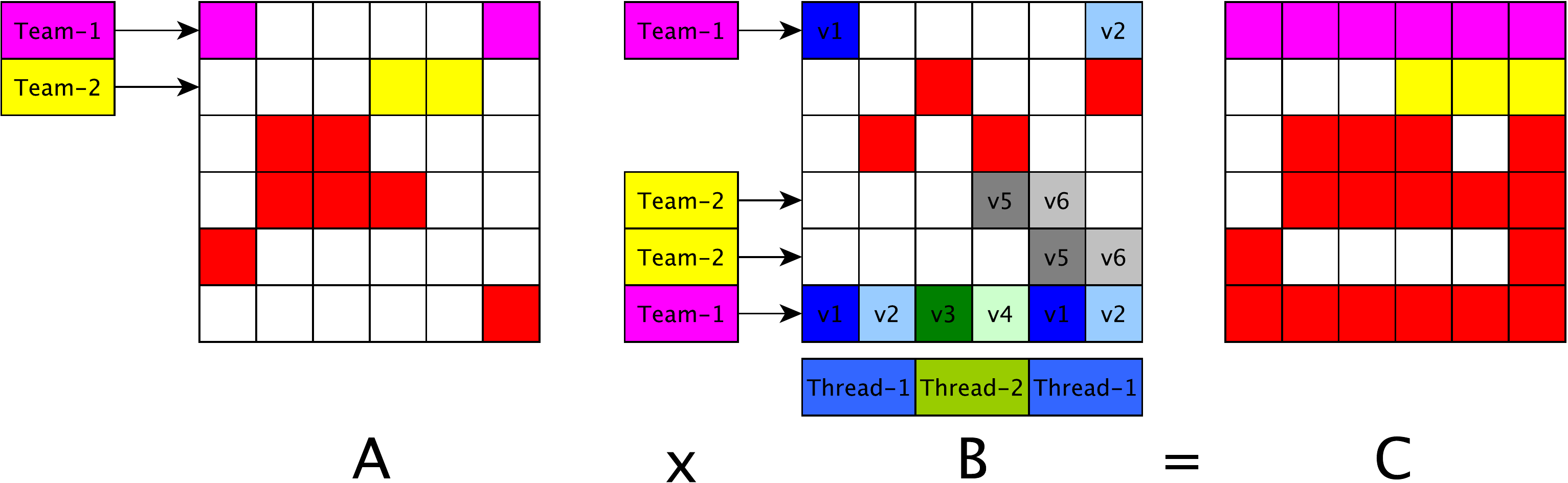}
    \label{fig:teamseq}}

    \caption{\extremelysmallfont Partitioning schemes for \spgemm{} using Kokkos-thread hierarchy. Nonzeroes and zeroes 
    are shown in red and white, respectively. Other colors represent the mapping of the data to execution units given 
    in Figure~\ref{fig:hier}.}
    \label{fig:spgemmhierarchy}
\end{figure}

\begin{figure}[t!]
    \centering
     \subfloat[\extremelysmallfont Thread-Parallel: Team-1 is assigned to a single row of $A$. 
				  Thread-1 and Thread-2 work on first and sixth rows of $B$ in parallel.
			          They further exploit vector parallelism for rows of $B$.]
    {\includegraphics[width=1\columnwidth]{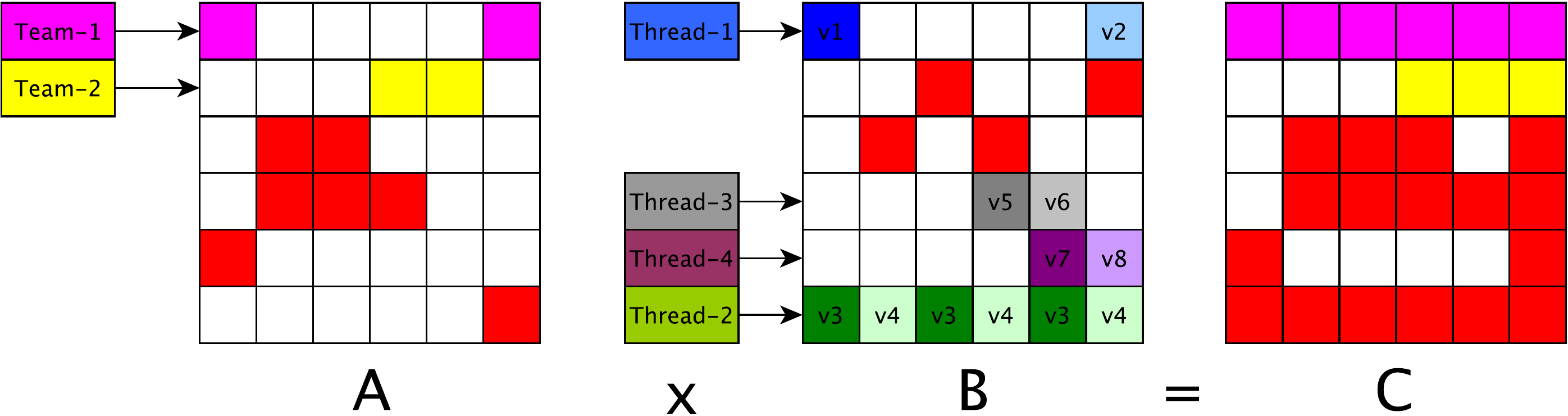}
    \label{fig:threadpar}}

     \subfloat[\extremelysmallfont Thread-Flat-Parallel: Team-1 is assigned to single row of $A$. 
				  The multiplications are flattened as shown in the bottom, and both 
				  thread and vector parallelism are exploited in this single dimension. 
				  Thread-1 and thread-2 work on different portions of the sixth row of $B$]
    {\includegraphics[width=1\columnwidth]{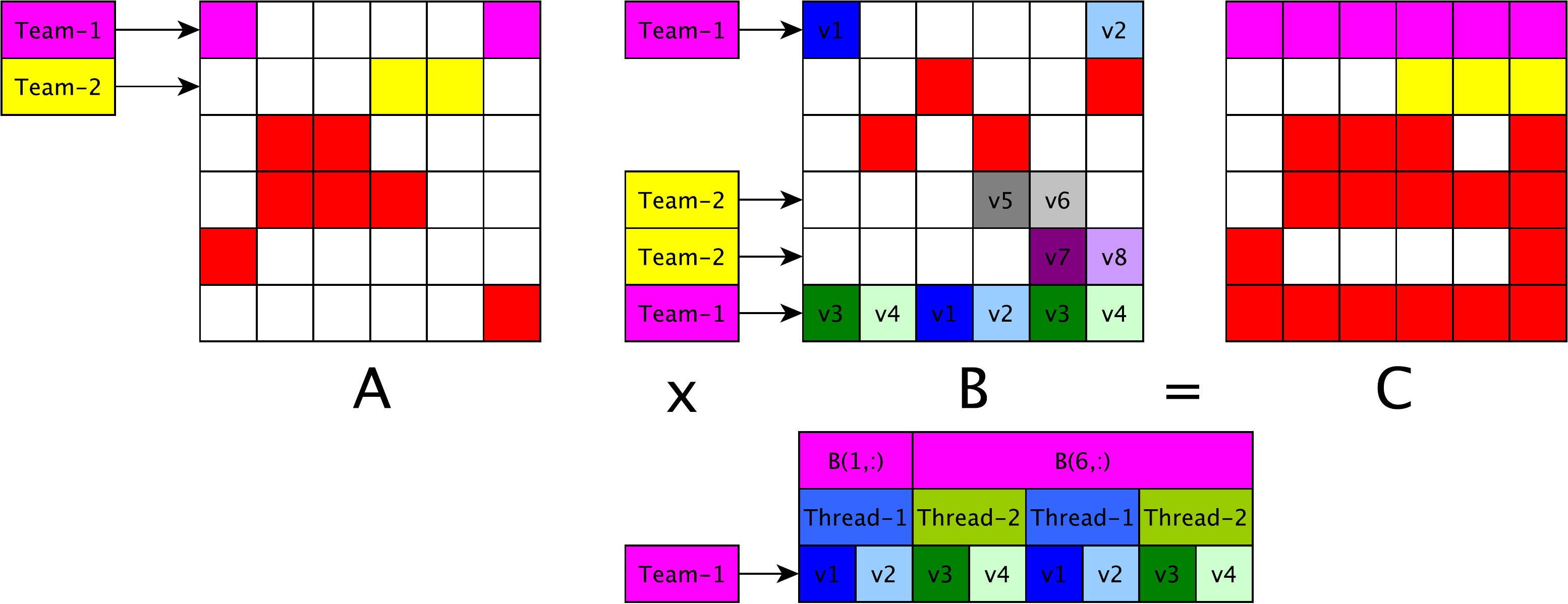}
    \label{fig:threadpar2}}

    \caption{\extremelysmallfont Partitioning schemes for \spgemm{} using Kokkos-thread hierarchy.}
    \label{fig:spgemmhierarchy2}
\end{figure}

Figure~\ref{fig:spgemmhierarchy} and ~\ref{fig:spgemmhierarchy2} give examples of different partitioning schemes.
Figure~\ref{fig:hier} shows our Kokkos-thread hierarchy used in this example.
%\todo{It would be nice if Figure 1 was upside down. Teams on top, split into
%thread etc}

%\begin{figure}[t!]
	
%\centering
%\includegraphics[width=1\columnwidth]{spgemm_partitioning}
%\caption{\verysmallfont
%\spgemm{} Kokkos thread hierarchy. Team-1 consists of $2$ Kokkos-threads, and 
%each Kokkos-thread has $4$ vector lanes (v[1-4]).
%}
%\label{fig:spgemmhierarchy}
%\end{figure}

\noindent
\textbf {Thread-Sequential:} %As shown in the example in Figure~\ref{fig:threadseq},
As shown in Figure~\ref{fig:threadseq},
this partitioning scheme assigns a group of rows to different teams, e.g. team-1 
gets the first two rows. Each thread within the team works on a different
row (Line-3 of Algorithm~\ref{alg:spgemmcoregeneral} is executed in parallel
by threads). Threads traverse the nonzeroes ($A(i,j)$) of their assigned row $A(i,:)$, and the 
corresponding rows $B(j,:)$ sequentially (Line-4). The nonzeroes of $B(j,:)$ 
are traversed, multiplied and inserted into accumulators using vector parallelism (Line-5). 
 A single thread computes the result for a single row of $C$ using vectorlanes.
%, multiplications of 
%the row are computed using different vector units. 
%Single Kokkos-thread computes a rows us
Our previous work,  {\sc kkmem}~\cite{deveci2017performance}, and AmgX 
follow this partitioning scheme.
Team resources (e.g., shared memory used for \Lone{} accumulators in \gpu{}s) are 
disjointly shared by the threads. This might cause more frequent use of \Ltwo{}
accumulators (located in slower memory space) for larger rows. 
The partitioning scheme, on the other hand, allows atomic-free accumulations. 
All computational units work on a single row of $B$ at a time, which
guarantees unique value insertions to the accumulators.
%\todo{May be a table w/ pros and cons of different partitioning scheme is a good idea}
%MD: I do not think we have any more space to add new things :(

\noindent
\textbf {Team-Sequential:} In Figure~\ref{fig:teamseq}, 
team-1 and team-2 are assigned the first and second row, respectively. Different from
Thread-Sequential, a whole team works on a single row of $A$ (Line-3 sequential).
Then, the whole team also sequentially traverses the nonzeroes ($A(i,j)$) of $A(i,:)$ (Line-4).
Finally, the nonzeroes in row $B(j,:)$ are traversed, multiplied and inserted into accumulators 
using both thread and vector parallelism (Line-5). 
%The method exploits both levels of 
%parallelism in the last loop.
This approach can use all of a team's resources when computing the result of a single row.
This allows \Lone{} to be larger, and thus reduces the number of \Ltwo{} accesses.
%The method uses the whole team resources for the calculation of the row, 
%allowing to have more space for \Lone{} accumulators and reducing the slower \Ltwo{}
%accesses. 
It also guarantees unique value insertions. 
%However, as the whole team works on a single row of $B$ sequentially, 
However, execution units are likely to be underutilized when the average 
row size of $B$ is small.
Unless we have a very dense multiplication, our preliminary experiments
show that this method does not have advantages over other methods. 
As a result, we do not use this method in our comparisons.

\noindent
\textbf {Thread-Parallel:} Figure~\ref{fig:threadpar} gives an example of this scheme.
This scheme assigns a whole team to a single row of $A$ (sequential Line-3). The method parallelizes 
both of the loops at Line-4 and Line-5. Threads are assigned to different nonzeroes of 
($A(i,j)$) of row $A(i,:)$, and the corresponding row $B(j,:)$. Nonzeroes in $B(j,:)$
are traversed, multipled and inserted into accumulators using vector parallelism (Line-5). 
%A row of $C$ is computed using whole team, while threads within the team are responsible from
%the multiplication of different nonzeroes of $A(i,j)$s with their corresponding $B(j,:)$.
As in Team-Sequential, more team resources are available for \Lone{}. 
The chance of underutilization is lower than in the previous method, but it can still happen
when rows require a very small number of multiplications.
%As different threads work on the rows of $B$ the chance of underutilization is lower
%than Team-Sequential.  Yet, it can suffer from underutilization when the row requires 
%less multiplications than the team's computation units. 
In addition, threads may suffer from load imbalance, when rows of $B$ differ in sizes.
%In addition, if one of the rows in $B$ is significantly larger than the other rows, as row-$6$ of Figure~\ref{fig:threadpar}.
%threads may suffer from load-balancing issues within the team. Lastly, 
This scheme does not guarantee unique insertions to accumulators, 
as different rows of $B$ are handled in parallel. This method is used in 
Nsparse~\cite{nagasaka2017high} and Kunchum {\it et al.}~\cite{kunchum2017improving}.

\noindent
\textbf {Thread-Flat-Parallel:} 
We use a Thread-Flat-Parallel scheme (Figure~\ref{fig:threadpar2}) to overcome
the limitations of the previous methods. This has also been explored 
%Dalton {\it et al.}~\cite{dalton2015optimizing} and Kunchum {\it et al.}
in~\cite{dalton2015optimizing} and~\cite{kunchum2017improving}. In this scheme,
a row of $A$ is assigned to a team, but as opposed to the Thread-Parallel scheme, 
this method flattens the second and third loop (Line-4 and Line-5).
%As before, the computation of a single row of $A$ is assigned to a single team.
%Different than Thread-Parallel, method flattens the second and third loop (Line-7 and Line-8)
%into a single loop. The loop size is the number of multiplications required for the row,
%and each vectorlane within threads is assigned single multiplication index. 
The single loop iterates over the total number of multiplications required for the row,
which is parallelized using both vector and thread parallelism.
Each vector unit calculates the index entries of $A$ and $B$ to work on, 
and inserts its multiplication result into the accumulators.
%They calculate calculate the index of the entries of $A$ and $B$ they will work on, 
%and insert the multiplication result into accumulator. 
This achieves a load-balanced distribution of the multiplications to 
execution units.
For example, both $B(1,:)$ and $B(6,:)$ are used for the multiplication of $A(1,:)$ in Figure~\ref{fig:threadpar2}.
Vectorlanes are assigned uniformly to the $8$ multiplications. In this scheme, a row of $B$ 
can be processed by multiple threads, and a single thread can work
accross multiple rows. Regardless of the differing row sizes in $B$, this method achieves
perfect load-balancing at the cost of performing index calculations for both $A$ and $B$.
%The row requires $8$ multiplications, and vectorlanes are assigned to different multiplications.
%Given a multiplication index, each vector multiplies corresponding nonzeroes in $A$ and $B$.
%A single row of $B$ can be processed by multiple threads. Similarly, vectorlanes of the single thread
%can work on different rows of $B$ at the same time. This allows a load-balanced work distribution
%regardless of the variety in the sizes of $B$ as in example.
The approach also provides larger shared memory for \Lone{} than Thread-Sequential. It may underutilize
compute units only when rows require a very small number of total multiplications. Parallel processing of the 
rows of $B$ does not guarantee unique insertions to accumulators.
%This scheme provides larger shared memory resources for \Lone{} accumulators than Thread-Sequential.
%It may suffer from underutilization only if the number of multiplications required for the row are 
%less than team's computation units. As different rows of $B$ are processed at a time, 
%partitioning do not guarantee unique insertions to accumulators. 

In this work, we use the Thread-Sequential and the Thread-Flat-Parallel scheme on \gpu{}s. 
These schemes behave similarly when teams have a single thread, our choice for \cpu{}s and \knl{}s. 
However, Thread-Flat-Parallel incurs index calculation overhead, which is not amortized
when there is not enough parallelism within a team. %As one obviously would expect, 
Thus, Thread-Sequential is used on \cpu{}s and \knl{}s. 
\subsubsection{Accumulators and Memory Pool Data Structures}
\label{sec:hashmap}
\myspace{-1.25ex}

%This section describes the different type of accumulators that supports parallel 
%insertions/accumulations and the memory pool data structure.
%Given the partitioning schemes above, we can describe the data structures to accumulate
%the entries of $C$.  
Our main methods use two-level, sparse hashmap-based accumulators.
Accumulators are used to compute the row size of $C$ in the symbolic phase, 
and the column indices and their values of $C$ in the numeric phase. 
Once teams/threads are created, they allocate some scratch memory 
(Line 1) for their private level-1 (\Lone{}) accumulator (not to be confused with the L1 cache).
This scratch memory maps to the \gpu{} shared memory in \gpu{}s and the default memory 
(i.e., {\sc ddr4} or high bandwidth memory) on \knl{}s. 
%The size of the accumulator depends on the size of a row of $C$.
If the \Lone{} accumulator runs out of space, global memory is allocated (Line 9)
in a scalable way using memory pools (explained below) for a \emph{row private} \Ltwo{} accumulator.
%This \Ltwo{} accumulator is used  for failed insertions from \Lone{}.
Its size is chosen to guarantee that it can hold all insertions.
%and its size is chosen to guarantee to hold all insertions. 
Upon the completion of a row computation, any allocated \Ltwo{} accumulator is explicitly released. 
Scratch spaces used by \Lone{} accumulators are automatically released by Kokkos when the threads retire. 
%\todo{I dont think if we need next sentence.}
%In order to implement the two-level, thread-scalable accumulator we
%need a thread-scalable memory pool (Section~\ref{sec:mempool}) which can be used by thread-scalable hash map
%(Section~\ref{sec:hashmap}).

We implemented three different types of accumulators. Two of these are sparse 
hashmap based accumulators, while the third one is a dense accumulator.

\noindent
\textbf{Linked List based HashMap Accumulator (LL):}
Accumulators are either thread or team {\em private} based on the partitioning scheme, so they need
to be highly scalable in terms of memory.
The hashmap accumulator here extends the hashmap used in~\cite{deveci2013hypergraph}
for parallel insertions. It consists of $4$ parallel arrays. Figure~\ref{fig:hashmap}
shows an example of a hashmap that has a capacity of $8$ hash entries and $5$ (key, value) pairs. 
The hashmap is implemented as a linked list structure. $Ids$ and $Values$ store the (key, value) pairs.
%which corresponds to column indices and their numerical values in numeric phase, 
%to {\sc csi} and {\sc cs} in symbolic phase. 
$Begins$ holds the beginning indices of the linked lists corresponding to the hash values, 
and $Nexts$ holds the indices of the next elements within the linked list.
For example, the set of keys that have a hash value of $4$ are stored with 
a linked list. The first index of this linked list is stored at $Begins[4]$. We use
this index to retrieve the (key, value) pairs ($Ids[0]$, $Values[0]$). 
The linked list is traversed using the $Nexts$ array.
An index value $-1$ corresponds to the end of the linked list for the hash value.
We choose the size of $Begins$ to be a power of $2$, therefore hash values can be calculated 
using {\sc BitwiseAnd}, instead of slow modulo ($\%$) operation. Each vector lane calculates the hash 
values, and traverses the corresponding linked list. If a key already exists in the hashmap,
values are accumulated.% with ``add'' and {\sc BitwiseOr} in numeric and symbolic phases, respectively.
%That is, a single row of $B$ is processed at a time, and the columns (keys) in the given row
%are unique. 
The implementation assumes that no values with the same keys are inserted concurrently.
%When implementing this method, we exploit the assumption that insertions at a time are 
%duplicate-free. This assumption holds for Thread-Sequential and Team-Sequential 
%partitioning schemes explained above.
%Thus, it avoids atomic operations for accumulations.
%are not needed for accumulation. 
If the key does not exist in the hashmap, vector lanes reserve the next available slot
with an atomic counter, and insert it to the beginning of the linked 
list (atomic\_compare\_and\_swap) of the corresponding hash.
%They set the $Begins$ of the corresponding hash
%value to their insertion index with atomic\_compare\_and\_swap, and the $Next$ of the 
%inserted index is set to the old beginning index. 
If it runs out out memory, it returns ``FULL'' and the failed insertions 
are accumulated in \Ltwo{}.
Because of its linked list structure, its performance is not affected by the 
occupancy of the hashmap. Even when it is full, extra comparisons are performed 
only for hash collisions. This provides constant complexity  
not only for \Lone{} but also for \Ltwo{} insertions, which are performed only 
when \Lone{} is full. 
This method assumes that concurrent insertions are duplicate-free and avoids
atomic operations for accumulations, which holds for Thread-Sequential and Team-Sequential.
When this assumption does not hold 
(Thread-Parallel and Thread-Flat-Parallel), a more complex implementation
with reduced performance is necessary. 

\noindent
\textbf{Linear Probing HashMap Accumulator (LP):}
Linear probing is a common technique that is used for hashing in the literature.
Nsparse applies this method for \spgemm{}. Figure~\ref{fig:linp} 
gives the example of a hashmap using LP. The data structure consists of two 
parallel arrays ($Ids$, $Values$). Initially each hash entry is set to $-1$ to 
indicate that it is empty.
Given an ($id$, $value$) pair, LP calculates a hash value and attempts to insert the pair into the hash location.
If the slot is taken, it performs a linear scan starting at the hash location and inserts it to the 
first available space.
For example, in Figure~\ref{fig:linp} hash for $28$ is calculated 
as $4$, but as the slot is taken it is inserted to the 
next available space. The implementation is straightforward and LP can easily be used 
with any of the $4$ partitioning schemes. However, as the occupancy of the hashmap 
becomes close to full, the hash lookups become very expensive. This makes it difficult 
to use LP in a two-level hashing approach. Each insertion to \Ltwo{} would first perform a full scan of 
\Lone{}, resulting in a complexity of $O(\|$\Lone{}$\|)$. 
Nsparse uses single-level LP, and when rows do not fit into \gpu{}s shared memory, 
this accumulator is directly allocated in global memory. 
In order to overcome this, we introduce a max occupancy parameter.
If the occupancy of \Lone{} is larger than this cut-off, we do not insert any new
$Ids$ to \Lone{} and use \Ltwo{} for failed insertions. We observe
significant slowdowns with LP once occupancy is higher than $50\%$, which
is used as a max occupancy ratio.

\noindent
\textbf{Dense Accumulators:} 
This approach accumulates rows in their dense format, requiring space 
$O(k)$ per thread/team. A dense structure allows columns 
to be accessed simply using their indices. This removes some overheads
such as hash calculation and collisions. Its implementation usually 
requires $2$ parallel arrays. The first is used for floating point 
values (initialized with $0$s). A second boolean array acts as a marker 
array to check if a column index was inserted previously. The column 
array of $C$ is used to hold the indices that are inserted in the dense
accumulator (requires an extra array in the symbolic phase). 
This is a single-level accumulator, and because of its high memory requirements 
dense accumulators are not suitable for \gpu{}s.
The approach is used by Patwary et al.~\cite{patwary2015parallel} with a column-wise 
partitioning of $B$ to reduce this memory overhead.

%\myspace{-3ex}
%\subsection{Memory Pool}
%\label{sec:mempool}
%\myspace{-1.25ex}
\noindent
{\textbf{Memory Pool:}} Algorithm~\ref{alg:spgemmcoregeneral} requires a portable, thread-scalable 
memory pool to allocate memory for \Ltwo{} ``sparse'' accumulators, in case a 
row of $C$ cannot fit into the \Lone{} accumulator.
The memory pool is allocated and initialized before the kernel call and 
services requests to allocate and release memory from thousands of threads/teams. 
As a result, allocate and release have to be thread scalable. 
Its allocate function returns a chunk of memory to a requestor thread and locks it. This 
lock is released as soon as the thread releases the chunk back to the pool.
The memory pool reserves {\sc numChunks} memory chunks, 
where each has a fixed size ({\sc chunkSize}). 
{\sc chunkSize} is chosen based on the ``maximum row size in $C$'' 
(\maxrowsize{}) to guarantee enough space for the work in any row of $C$. 
\maxrowsize{} is not known before performing the symbolic phase so it uses an upper bound.
The upper bound is the maximum number of multiplies
(\maxrowflops{}) required by any row. 
That number can be computed by summing the size of all rows of $B$ that contribute to a row. 
The memory pool has two operational modes: unique and non-unique mapping of 
chunks to threads ({\sc one2one} and {\sc many2many}). 

The parameters of the memory pool are architecture specific.
{\sc numChunks} is chosen based on the available concurrency in an architecture. 
It is an exact match to the number of threads on the \knl{}s/\cpu{}s. On \gpu{}s, 
we over-estimate the concurrency to
efficiently acquire memory. We check the available memory, and reduce {\sc numChunks} if the memory 
allocation becomes too expensive on \gpu{}s. \cpu{}s/\knl{}s use {\sc one2one} and \gpu{}s use 
{\sc many2many}. The allocate function of the memory pool uses thread indices. These indices assist 
the look-up for a free chunk. The pool directly returns the chunk with the given thread index when using the
{\sc one2one} mode. 
This allows \cpu{}/\knl{} threads to reuse local NUMA memory regions. 
In the {\sc many2many} mode, the pool starts a scan from the given thread-index until an available chunk is found. 
If the memory pool does not immediately have a memory chunk available to fulfill a request, the requesting 
computational unit spins until it successfully receives an allocation. 

\myspace{-3ex}
\subsection{Compression}
\label{sec:compression}
\myspace{-1.25ex}

\begin{figure}[t!]
   \centering
     \subfloat[Compression]
    {\includegraphics[width=0.30\columnwidth]{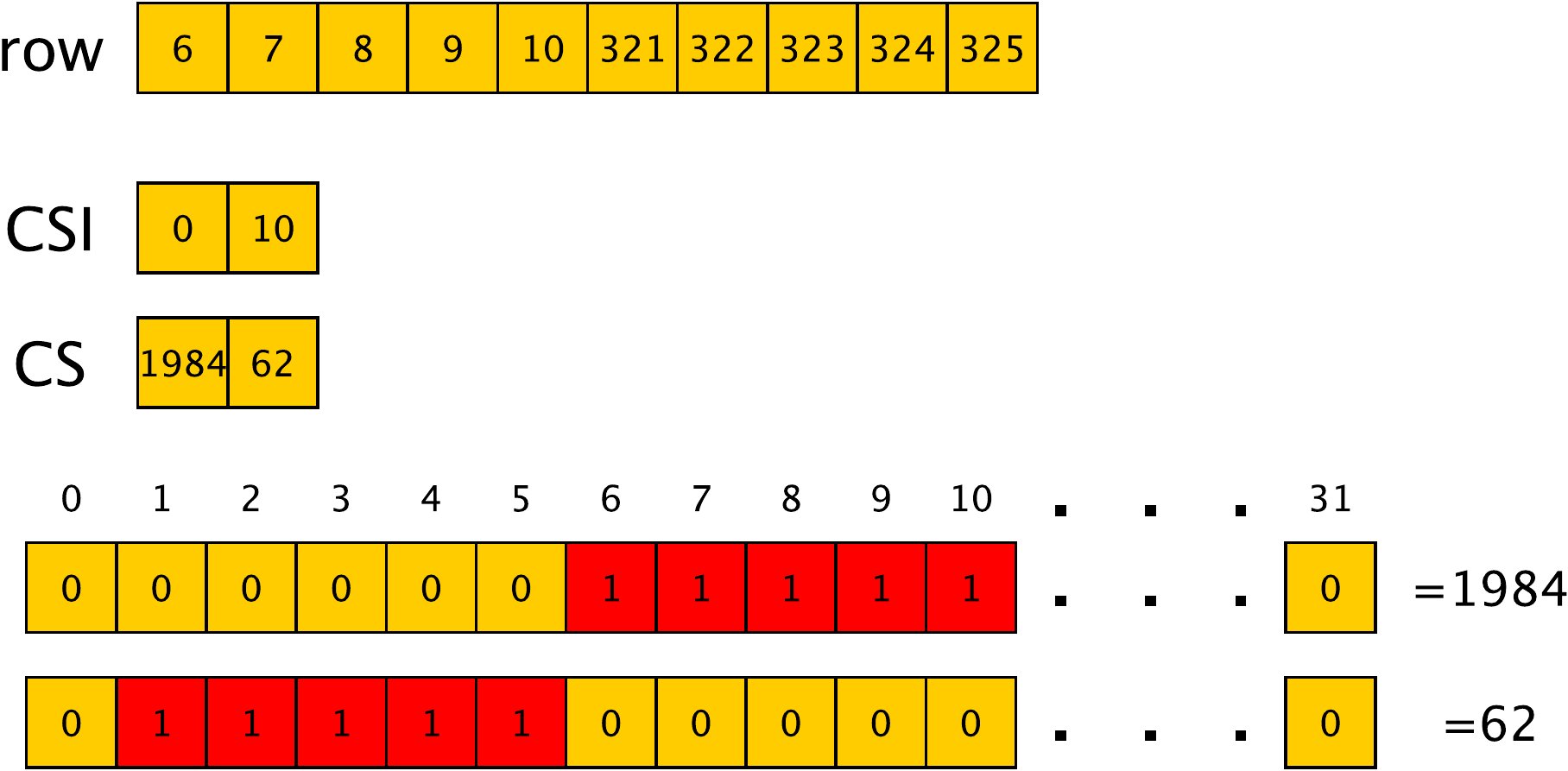}
    \label{fig:compresion}}
     \subfloat[LL]
    {\includegraphics[width=0.30\columnwidth]{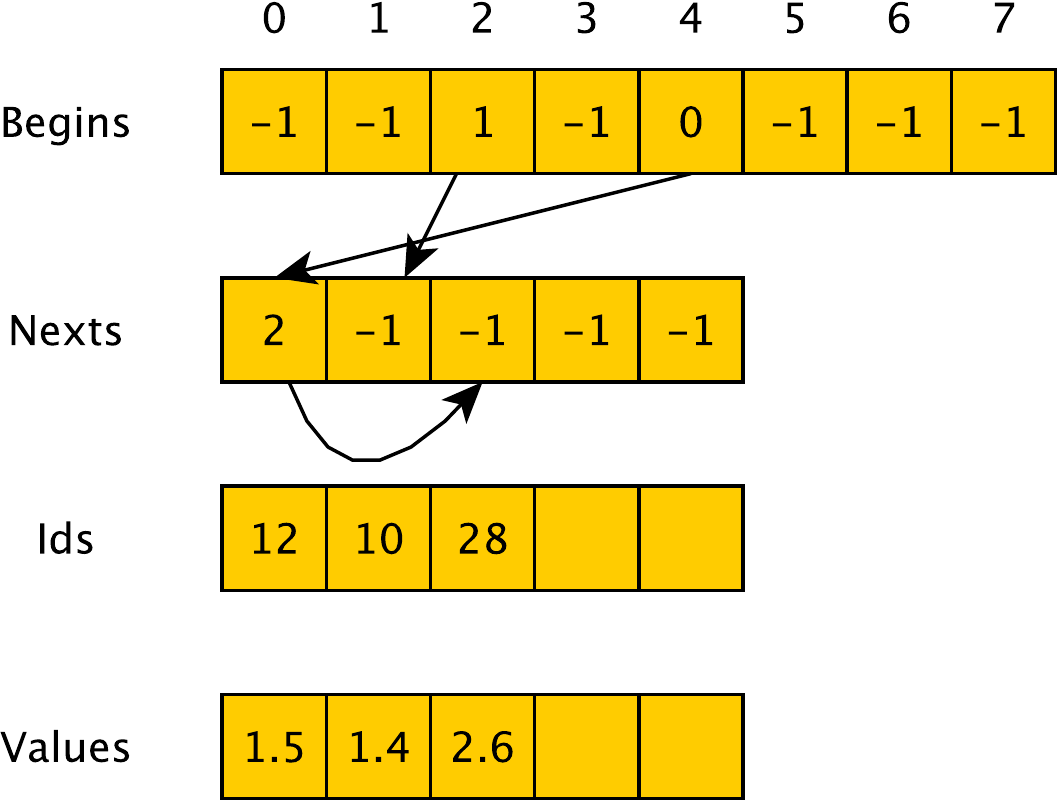}\label{fig:hashmap}}
     \subfloat[LP ]
    {\includegraphics[width=0.30\columnwidth]{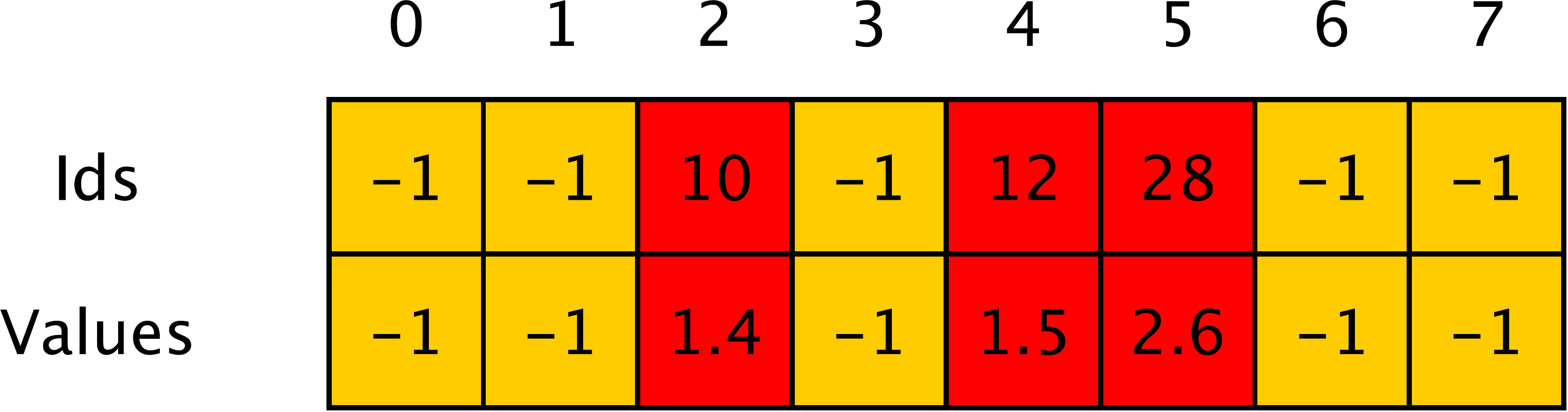}\label{fig:linp}}
    \caption{\verysmallfont Compression and Hashmap examples} 
\end{figure}

Compression is applied to $B$ in the symbolic phase.
%This section addresses the problem of compressing the graph of $B$ in the symbolic phase. 
This method, based on packing columns of $B$ as bits, can
reduce the size of $B$'s graph up to $32\times$ (the number of bits in an integer).
The graph structure of $B$ encodes binary relations -
existence of a nonzero in $(i,j)$ or not. 
This can be represented using single bits. 
We compress the rows of $B$ such that $32$ columns of $B$ 
are represented using a single integer following
the color compression idea in~\cite{deveci16coloring}. 
In this scheme, the traditional column index array in a compressed-row
matrix is represented with $2$ arrays of smaller size: 
``column set'' ({\sc cs})  and ``column set index'' ({\sc csi}). 
Set bits in {\sc cs} denote existing columns. That is, if 
the $i^{th}$ bit in {\sc cs} is $1$, the row has a nonzero entry at the 
$i^{th}$ column. {\sc cs} is used to represent more than $32$ columns. 
Figure~\ref{fig:compresion} shows an example of the compression of a row with $10$ columns.
\mdcomment {The original symbolic phase would insert all $10$ columns for this row
into accumulators. Compression reduces the row size, and only $2$ 
are inserted into an accumulator with {\sc BitwiseOr} operation on {\sc cs} values.}
It is more successful if the column indices in each row are packed 
close to each other.

%%%%%%%%%%%%MD below should go in arxiv version
In Algorithm~\ref{alg:spgemmcoregeneral}, which computational units scan the 
rows of $A$ and $C$ only once.
However, a nonzero value in $B$ is read multiple times, and 
there are $flops$ accesses to $B$;
i.e., $B(i,:)$ is read 
as many times as the size of $A(:,i)$. 
Assuming uniform structure for $A$
with $\delta_{A}$ (average degree of $A$)  
nonzeroes in each column, each
row of $B$ is accessed $\delta_{A}$ times. Thus, \flops{} becomes $O(\delta_{A} \times nnz_{B})$,
where $nnz_{B}$ is the number of nonzeroes in $B$. 
If a compression method with linear time complexity ($O(nnz_{B})$), as the one
above, reduces the size 
of $B$ by some ratio $CF$, the amount of work in the symbolic can be reduced by 
$O(CF \times \delta_{A} \times nnz_{B})$. %, with a compression cost of $O(nnz_{B})$. 

Compression reduces the problem size, allows faster row-union operations 
using {\sc BitwiseOr}, and makes the symbolic phase more efficient.
The reduction in row lengths of $B$ also reduces the calculated \maxrowflops{}, %$max\_row\_flops$, 
the upper bound prediction for the required memory of accumulators in the symbolic 
phase, further improving the robustness and scalability of the method. 
%The effectiveness of compression is shown .
%For example, the size of the dense accumulators are reduced from $k$ to $\frac{k}{32}$.
However, compression is not always successful at reducing the matrix size. 
For matrices in which column indices are spread, the compression may not reduce 
the input size, and introduction of the extra values ({\sc cs}) may slow
the symbolic phase down. For this reason, we run compression in two phases. We first
calculate the row sizes in the compressed matrix, and calculate the overall 
\flops{} after the compression. If \flops{} is reduced more than $15\%$, 
the matrix is compressed and the symbolic phase is executed using this compressed matrix.
Otherwise, we do not perform compression and run the symbolic phase using the original matrices.
We find this compression method to be very effective in practice; e.g., the \flops{} reduction
is less than $15\%$ only for $7$ of $83$ test cases used in this paper.
See~\cite{wolf2017fast} for the effect of this compression method on solving the triangle
counting problem.

%There exist more complicated graph compression methods in the literature. 
%Some uses delta encoding to compress the columns of a row~\cite{%blandford2004experimental, 
%willcock2006accelerating}, while others seek for similarity of the rows (block structures)
%\cite{%adler2001towards, 
%gilbert2004compressing, kourtis2011csx, deveci2013hypergraph}. 
%These methods are rather expensive, and it is can be amortized on repetitive or computationally
%intensive problems. 
%We employ compression for symbolic phase (executed only once), none 
%of these methods are likely to amortize. The similarity based ones could be used to improve
%the quality of the compression.
%We avoid using such expensive methods, and evaluate the effectiveness of this light-weight compression 
%in the experiments. Also, see~\cite{wolf2017fast} for the effect of this compression method on real
%graph analytics problems. 

\myspace{-3ex}
\subsection{KokkosKernels {\sc SpGEMM} Methods}
\myspace{-1.25ex}

Our previous work~\cite{deveci2017performance} proposes the {\sc kkmem} algorithm.
It uses a Thread-Sequential approach with LL accumulators.
Its auto parameter detection focuses on the selection of vector-length.
This size is fixed for all threads in a parallel kernel. 
We set it on \gpu{}s by rounding %the average number of nonzeroes in a row of $B$ 
$\delta_{B}$ (average degree of $B$) to the closest power of $2$ (bounded by warp size $32$). 
On \knl{}s and \cpu{}s, Kokkos sets the length depending on the compiler and underlying architecture 
specifications. 
The size of its \Lone{} accumulators depends on the available 
\emph{shared memory} on \gpu{}s. 
The size of the \Ltwo{} accumulator (in the \emph{global memory})
is chosen as \maxrowsize{} in the numeric (\maxrowflops{} in the symbolic).
%The upper bound for the numeric phase is 
%the maximum of the actual row sizes ($max\_row\_size$) as the actual sizes 
%are known. 
In contrast to \gpu{}s, both \Lone{} and \Ltwo{} accumulators are in 
the same memory space on \knl{}s/\cpu{}s.
Since there are more resources per thread on the \knl{}s/\cpu{}s, %we skip L1 accumulator, and only 
we make \Lone{} big enough to hold \maxrowsize{} (or \maxrowflops{}).
This is usually small enough to fit into cache on \knl{}s/\cpu{}s. 

{\sc kkmem} is designed to be scalable to run on large datasets with large thread counts. 
It aims to minimize the memory use ($O($\maxrowsize{}$)$) and to localize memory accesses at the 
cost of increased hash operations/collisions. 
In this work, we add {\sc kkdense} that uses dense accumulators ($O(k)$) and runs only on 
\cpu{}s and \knl{}s. It does not have the extra cost of hash operations.
However, its memory accesses may not be localized depending on the structure of a problem.
When $k$ is small, using sparse accumulators does not have much 
advantage over dense accumulators (on \knl{}s/\cpu{}s) as a dense accumulator would 
also fit into cache. Moreover, some matrix multiplications might result in
\maxrowsize{} to be very close to $k$ (e.g. squaring RMAT matrices results in 
\maxrowsize{} to be $95\%$ of $k$). In such cases sparse accumulators allocate as 
much memory as dense accumulators, while still performing extra hash operations.
%In these multiplications, a scalable \spgemm{} method using sparse accumulators
%is naturally not expected to perform better than a simpler method using dense accumulators. 
Sparse accumulators are naturally not expected to perform better than dense accumulators for these cases. 

This work proposes a meta algorithm {\sc kkspgemm} that chooses either of these methods
on \cpu{}s and \knl{}s based on the size of $k$. We observe superior performance of {\sc kkdense}
for $k < 250,000$ on \knl{}'s {\sc ddr} memory. As $k$ gets larger {\sc kkmem} outperforms 
{\sc kkdense}. We introduce a cut-off parameter for $k$ based on this observation. The meta-algorithm 
runs {\sc kkdense} for $k < 250,000$, and {\sc kkmem} otherwise. As the columns are compressed 
in the symbolic phase by a factor of $32$, {\sc kkspgemm} may run {\sc kkdense} for the symbolic 
phase, and {\sc kkmem} for the numeric phase. 
A more sophisticated selection of this parameter requires consideration of the underlying architecture.
If the architecture has a larger memory bandwidth,
%, or if it is not using all available cores (not saturating the bandwidth), 
it may be more tolerant to larger dense accumulators.
For example, using {\sc mcdram} or cache-mode in \knl{}s provides larger memory bandwidth,
and {\sc kkdense} also achieves better performance than {\sc kkmem} for $k > 250,000$. 
Yet, in the rest of the paper we use $k=250,000$ as cut-off across different architectures,
which captures the best methods for most cases.

The parameter selection on \gpu{}s is more complicated with additional
variables, i.e., shared memory, warp (vector-length) and block sizes. 
%{\sc kkmem} uses Thread-Sequential partitioning with LL accumulators.
This work introduces {\sc kklp}, which uses the Thread-Flat-Parallel partitioning with 
two-level LP accumulators. For problems in which rows require few (on average $< 256$) 
multiplications, our meta algorithm runs {\sc kkmem}; otherwise it runs {\sc kklp}.
Once the algorithm is chosen, based on the {\em average output row size} (\avgrowsize{}), we adjust the 
shared memory size (initially $16$KB per team) to minimize the use of \Ltwo{} accumulators. 
For {\sc kkmem}, 
if \avgrowsize{} does not fit into \Lone{}, we reduce the number of threads 
within teams (by increasing the vector length up to $32$ and reducing threads 
at the same time) to increase the shared memory per thread, so that most of the entries can fit into \Lone{}.
For {\sc kklp}, if initial available shared memory for the team ($16$KB) 
provides more space than \avgrowsize{}, we reduce the team size and its 
shared memory to be able to run more blocks concurrently on the streaming multiprocessors of \gpu{}s.
If \avgrowsize{} requires larger memory than $16$KB, we increase the 
shared memory at most to $32$KB (and block size to $512$). 
As the row sizes are unknown at the beginning of the symbolic phase,
it is more challenging to select these parameters then. We estimate 
\avgrowsize{} from \flops{} by assuming every $n$th ($8$th is used for the experiments) 
multiplication will reduce to the same nonzero. 

%%%%%%%%%%%%%%%%%BELOW is for arxiv version
The experiments run our old method {\sc kkmem} without this parameter selection
to highlight the improvements w.r.t. previous work.
Table~\ref{tab:methods} summarizes the methods used in this paper.
Our implementations cannot launch concurrent kernels using 
cuda-streams as Nsparse does, as Kokkos does not support that yet. Instead, we launch 
a single kernel using the above parameter selection.

\begin{table}
\centering
\caption{\verysmallfont The KokkosKernels variants used in this paper.}
\label{tab:methods}
\resizebox{\columnwidth}{!}{%
\begin{tabular}{l|l|l|l}
\cpu{}s \& \knl{}s & {\sc kkmem}~\cite{deveci2017performance} &{\sc kkdense}: Dense Acc. & {\sc kkspgemm}: {\sc kkmem} for $k<250,000$, {\sc kkdense} otherwise. \\ \hline
\gpu{}s & {\sc kkmem}~\cite{deveci2017performance} & {\sc kklp}: 2-level LP with Thread-Flat-Parallel & {\sc kkspgemm}: {\sc kkmem} for average row flops$<256$, {\sc kklp} otherwise. \\ 
\end{tabular}
}
\end{table}
\section{Experiments} 
\label{sec:results}
\myspace{-2.5ex}

\begin{table}[]
\centering
\caption{\verysmallfont The specifications of the architectures used in the experiments.
The experiments set {\tt OMP\_PROC\_BIND=spread} and {\tt OMP\_PLACES=threads}.}
\label{tab:arch}
\resizebox{\columnwidth}{!}{%
\begin{tabular}{l|l|l|l}
Cluster - \cpu{}/\gpu{} & Bowman - Intel KNL & White (Host) - IBM Power8 & White (GPU) - NVIDIA P100-SXM2\\ \hline
%\cpu{}/GPU & Intel KNL & IBM Power8 & NVIDIA P100-SXM2\\ \hline
Compiler & intel 18.0.128 & gnu 5.4.0  & gnu 5.4.0, nvcc 8.0.44\\ \hline
Core specs &$68\times1.40$GHz cores, 4 hyperthreads %\begin{tabular}[c]{@{}l@{}}\end{tabular} 
           & $16\times 3.60$ GHz cores, 8 hyperthreads %\begin{tabular}[c]{@{}l@{}}\end{tabular}  
           & $1.48$GHz  %\begin{tabular}[c]{@{}l@{}}Com. Cap. \\ 6.0\end{tabular} 
           \\ \hline
Memory & $16$ GB MCDRAM $460$ GB/s, $96$ GB DDR4 $102$ GB/s %\begin{tabular}[c]{@{}l@{}}- $16$ GB MCDRAM \\ $460$ GB/s \\ - $96$ GB DDR4 \\ $102$ GB/s\end{tabular} 
       & $512$ GB DDR4, $2$ NUMA %\begin{tabular}[c]{@{}l@{}}$512$ Gb \\ $2$ NUMA\end{tabular}
       & $16$ GB HBM\\ %\hline
\end{tabular}

%\begin{tabular}{l|l|l|l}
%Cluster & Bowman & White (Host) & White (GPU) \\ \hline
%\cpu{}/GPU & KNL & Power8 & P100 \\ \hline
%Compiler & icc 18.0.128 & \begin{tabular}[c]{@{}l@{}}gcc 5.4.0\end{tabular} & \begin{tabular}[c]{@{}l@{}}gcc 5.4.0\\ cuda 8.0.44\end{tabular} \\ \hline
%Core specs &\begin{tabular}[c]{@{}l@{}}$68\times1.40$GHz \\ cores \\ 4 hyperthreads\end{tabular} 
%           & \begin{tabular}[c]{@{}l@{}}$16\times 3.60$ GHz \\ cores \\ 8 hyperthreads\end{tabular}  
%           & \begin{tabular}[c]{@{}l@{}}Com. Cap. \\ 6.0\end{tabular} \\ \hline
%Memory & \begin{tabular}[c]{@{}l@{}}- $16$ GB MCDRAM \\ $460$ GB/s \\ - $96$ GB DDR4 \\ $102$ GB/s\end{tabular} 
%       & \begin{tabular}[c]{@{}l@{}}$512$ Gb \\ $2$ NUMA\end{tabular}
%       & $16$ GB \\ \hline
%\end{tabular}
}
\end{table}

%\subsection{Setup} 
Performance experiments are performed on three different configurations, representing 
two of the most commonly HPC leadership class machine hardware designs:
Intel XeonPhi and IBM Power with NVIDIA GPUs.  
The configurations of the nodes are listed in Table~\ref{tab:arch}.
Our methods are implemented using the Kokkos library (2.5.00), and will be 
available in KokkosKernels (2.5.10). Detailed explanation about the raw experiment results and 
reproducing them can be found at \url{https://github.com/kokkos/kokkos-kernels/wiki/SpGEMM\_Benchmarks}.
Each run reported in this paper is the average of 
$5$ executions (preceded with $1$ excluded warmup run) with double precision arithmetic and $32$ bit integers.
We evaluate $83$ matrix multiplications, $24$ of which are of the form 
$R \times A \times P$ as found in multigrid, while the rest are
of the form $A \times A$ using matrices from the UF sparse suite~\cite{davis2011university}.
The problems are listed in Table~\ref{tab:overall1}.% and~\ref{tab:overall2}.
\mdcomment{Experiments are run for both a NoReuse and a Reuse case.  
Both the symbolic and the numeric phase are executed for NoReuse. Reuse executes only 
the numeric phase, and reuses the previous symbolic computations. Unless specifically stated, 
the results refer to the NoReuse case.}

%The experiments are organized in four parts.  Section~\ref{sec:cpuexp},
%Section~\ref{sec:knlexp} and Section~\ref{sec:gpuexp} evaluate the performance of the methods 
%on Power8, \knl{}s and \gpu{}s respectively. Section~\ref{sec:compressionexp} evaluates 
%the effect of the compression method. %Overall results are summarized 
%in Section~\ref{sec:overallexp}.

\begin{table}[]
\centering
\caption{The list of the matrices used in the experiments in this paper.
CF and CMRF gives the ratio of the reduction in overall number of flops 
and maximum row flops. Last four columns list the achieved GFLOPs/sec
by {\sc kkspgemm} on $4$ architectures.}
\label{tab:overall1}
\resizebox{\columnwidth}{!}{

\begin{tabular}{r|r|r|r|r|r|r|r|r|r|r||r|r|r|r}
ID & Multiplication & $m$ & $n$ & $k$ & \flops{} & \maxrowflops{} &$\|$C$\|$ & \maxrowsize & CF & CMRF &  Power8 & P100 & \begin{tabular}[c]{@{}r@{}}KNL\\ DDR\end{tabular} & \begin{tabular}[c]{@{}r@{}}KNL\\ CACHE\end{tabular} \\ 
\hline
1 & amazon0302 & 262,111 & 262,111 & 262,111 & 6,021,131 & 25 & 3,896,236 & 25 & 0.71 & 1.00 & 1.63 & 1.64 & 0.80 & 0.89 \\ \hline
2 & belgium\_osm & 1,441,295 & 1,441,295 & 1,441,295 & 7,017,228 & 25 & 5,323,073 & 18 & 0.65 & 0.80 & 1.36 & 1.13 & 0.60 & 0.66 \\ \hline
3 & mac\_econ\_fwd500 & 206,500 & 206,500 & 206,500 & 7,556,897 & 229 & 6,704,899 & 215 & 0.57 & 0.59 & 2.22 & 1.49 & 0.51 & 0.72 \\ \hline
4 & mc2depi & 525,825 & 525,825 & 525,825 & 8,391,680 & 16 & 5,245,952 & 10 & 0.76 & 0.94 & 2.55 & 1.51 & 1.00 & 1.65 \\ \hline
5 & delaunay\_n18 & 262,144 & 262,144 & 262,144 & 9,907,810 & 214 & 5,430,294 & 154 & 0.47 & 0.62 & 3.25 & 2.18 & 1.07 & 1.38 \\ \hline
6 & 2cubes\_sphere & 101,492 & 101,492 & 101,492 & 27,450,606 & 544 & 8,974,526 & 180 & 0.48 & 0.63 & 5.15 & 5.80 & 1.45 & 2.57 \\ \hline
7 & ca-HepPh & 12,008 & 12,008 & 12,008 & 30,793,000 & 93,923 & 3,284,660 & 3,211 & 0.73 & 0.03 & 7.59 & 1.48 & 2.36 & 2.35 \\ \hline
8 & rgg\_n\_2\_18\_s0 & 262,144 & 262,144 & 262,144 & 39,648,378 & 716 & 9,179,295 & 67 & 0.81 & 0.70 & 5.09 & 6.82 & 2.51 & 3.20 \\ \hline
9 & hugetrace-00000 & 4,588,484 & 4,588,484 & 4,588,484 & 41,260,426 & 9 & 28,308,760 & 7 & 0.74 & 1.00 & 1.06 & 3.16 & 0.54 & 0.83 \\ \hline
10 & web-Stanford & 281,903 & 281,903 & 281,903 & 44,110,669 & 13,682 & 20,811,442 & 3,421 & 1.00 & 0.64 & 3.60 & 2.12 & 1.69 & 2.13 \\ \hline
11 & Stanford & 281,903 & 281,903 & 281,903 & 44,110,669 & 491,041 & 20,811,442 & 68,455 & 0.86 & 0.03 & 1.99 & 0.19 & 0.44 & 0.83 \\ \hline
12 & amazon-2008 & 735,323 & 735,323 & 735,323 & 46,082,867 & 100 & 25,366,745 & 100 & 0.38 & 0.93 & 3.88 & 3.65 & 2.03 & 2.67 \\ \hline
13 & web-Google & 916,428 & 916,428 & 916,428 & 60,687,836 & 4,334 & 29,710,164 & 2,256 & 1.00 & 1.00 & 2.13 & 1.81 & 1.21 & 1.92 \\ \hline
14 & webbase-1M & 1,000,005 & 1,000,005 & 1,000,005 & 69,524,195 & 116,179 & 51,111,996 & 12,383 & 0.25 & 0.27 & 4.23 & 1.19 & 1.54 & 1.82 \\ \hline
15 & offshore & 259,789 & 259,789 & 259,789 & 71,342,515 & 562 & 23,356,245 & 182 & 0.61 & 0.68 & 5.22 & 7.29 & 2.75 & 3.97 \\ \hline
16 & conf5\_4-8x8-05 & 49,152 & 49,152 & 49,152 & 74,760,192 & 1,521 & 10,911,744 & 222 & 0.21 & 0.26 & 12.34 & 10.78 & 4.42 & 5.45 \\ \hline
17 & delaunay\_n21 & 2,097,152 & 2,097,152 & 2,097,152 & 79,241,506 & 219 & 43,417,524 & 157 & 0.47 & 0.65 & 3.55 & 4.95 & 1.73 & 2.48 \\ \hline
18 & cop20k\_A & 121,192 & 121,192 & 121,192 & 79,883,385 & 2,489 & 18,705,069 & 495 & 0.39 & 0.43 & 6.90 & 10.11 & 2.94 & 4.40 \\ \hline
19 & cit-Patents & 3,774,768 & 3,774,768 & 3,757,431 & 82,152,992 & 6,142 & 68,848,721 & 3,925 & 0.99 & 0.99 & 0.89 & 0.90 & 0.54 & 1.07 \\ \hline
20 & filter3D & 106,437 & 106,437 & 106,437 & 85,957,185 & 3,340 & 20,161,619 & 550 & 0.54 & 0.63 & 7.61 & 8.16 & 3.29 & 5.11 \\ \hline
21 & Empire\_RAxP & 8,800 & 2,160,000 & 8,800 & 91,604,280 & 11,037 & 280,800 & 36 & 0.59 & 0.03 & 5.86 & 3.02 & 2.78 & 3.61 \\ \hline
22 & Empire\_RxAP & 8,800 & 2,160,000 & 8,800 & 91,604,280 & 11,064 & 280,800 & 36 & 0.39 & 0.03 & 6.24 & 5.12 & 3.17 & 4.38 \\ \hline
23 & cnr-2000 & 325,557 & 325,557 & 325,557 & 96,065,788 & 34,537 & 34,174,066 & 15,723 & 0.18 & 0.19 & 7.32 & 2.13 & 3.16 & 5.12 \\ \hline
24 & soc-Slashdot0811 & 77,360 & 77,360 & 77,360 & 111,839,175 & 134,321 & 78,851,659 & 31,750 & 0.60 & 0.03 & 4.58 & 1.56 & 1.46 & 2.46 \\ \hline
25 & amazon0601 & 403,394 & 403,394 & 403,394 & 149,306,190 & 31,313 & 98,600,816 & 20,607 & 0.79 & 0.40 & 3.40 & 1.40 & 1.20 & 2.03 \\ \hline
26 & rma10 & 46,835 & 46,835 & 46,835 & 156,480,259 & 12,765 & 7,900,917 & 425 & 0.10 & 0.10 & 15.89 & 16.65 & 8.25 & 9.51 \\ \hline
27 & hugebubbles-00000 & 18,318,143 & 18,318,143 & 18,318,143 & 164,791,952 & 9 & 113,009,849 & 7 & 0.74 & 1.00 & 1.03 & 4.02 & 0.73 & 1.02 \\ \hline
28 & hugebubbles-00020 & 21,198,119 & 21,198,119 & 21,198,119 & 190,713,076 & 9 & 132,690,161 & 7 & 0.83 & 1.00 & 0.57 & 3.28 & 0.57 & 0.82 \\ \hline
29 & rgg\_n\_2\_20\_s0 & 1,048,576 & 1,048,576 & 1,048,576 & 194,980,566 & 1,011 & 41,709,507 & 75 & 0.90 & 0.74 & 5.28 & 8.10 & 3.37 & 4.24 \\ \hline
30 & Elasticity\_113\_RxAP & 54,872 & 4,328,691 & 54,872 & 205,253,787 & 3,993 & 1,404,928 & 27 & 0.47 & 0.43 & 5.25 & 5.75 & 3.13 & 4.47 \\ \hline
31 & Elasticity\_113\_RAxP & 54,872 & 4,328,691 & 54,872 & 205,253,787 & 3,993 & 1,404,928 & 27 & 0.65 & 0.43 & 5.25 & 3.51 & 2.75 & 4.11 \\ \hline
32 & Stanford\_Berkeley & 683,446 & 683,446 & 683,446 & 222,116,841 & 955,984 & 78,130,972 & 136,877 & 0.17 & 0.03 & 4.49 & 0.49 & 1.40 & 1.57 \\ \hline
33 & europe\_osm & 50,912,018 & 50,912,018 & 50,912,018 & 241,277,568 & 44 & 182,570,158 & 28 & 0.64 & 0.80 & 1.32 & 2.82 & 0.87 & 1.17 \\ \hline
34 & cant & 62,451 & 62,451 & 62,451 & 269,486,473 & 5,913 & 17,440,029 & 375 & 0.12 & 0.12 & 15.82 & 18.99 & 9.36 & 12.89 \\ \hline
35 & Brick\_185\_RxAP & 238,328 & 6,331,625 & 238,328 & 307,568,462 & 2,220 & 6,436,594 & 78 & 0.47 & 0.71 & 5.37 & 6.79 & 3.02 & 4.21 \\ \hline
36 & Brick\_185\_RAxP & 238,328 & 6,331,625 & 238,328 & 307,568,462 & 2,217 & 6,436,594 & 78 & 0.64 & 0.78 & 5.25 & 3.82 & 3.07 & 4.16 \\ \hline
37 & BigStar\_4657\_RAxP & 1,446,620 & 21,687,649 & 1,446,620 & 369,829,182 & 265 & 24,552,796 & 17 & 0.79 & 0.80 & 2.47 & 4.53 & 1.36 & 2.63 \\ \hline
38 & BigStar\_4657\_RxAP & 1,446,620 & 21,687,649 & 1,446,620 & 369,829,182 & 265 & 24,552,796 & 17 & 0.63 & 0.67 & 2.29 & 5.64 & 1.27 & 2.68 \\ \hline
39 & shipsec1 & 140,874 & 140,874 & 140,874 & 450,639,288 & 6,876 & 24,086,412 & 342 & 0.15 & 0.17 & 15.67 & 21.58 & 10.63 & 13.30 \\ \hline
40 & consph & 83,334 & 83,334 & 83,334 & 463,845,030 & 6,561 & 26,539,736 & 375 & 0.15 & 0.24 & 16.10 & 25.88 & 11.03 & 13.72 \\ \hline
41 & cage14 & 1,505,785 & 1,505,785 & 1,505,785 & 532,205,737 & 1,525 & 236,999,813 & 646 & 0.78 & 0.74 & 4.76 & 4.85 & 2.89 & 4.11 \\ \hline
42 & pdb1HYS & 36,417 & 36,417 & 36,417 & 555,322,659 & 32,222 & 19,594,581 & 987 & 0.07 & 0.04 & 18.64 & 24.47 & 12.69 & 15.60 \\ \hline
%\end{tabular}}
%\end{table}

%\begin{table}[]
%\centering
%\caption{Second part of Table~\ref{tab:overall1}.}
%\label{tab:overall2}
%\resizebox{\columnwidth}{!}{
%\begin{tabular}{r|r|r|r|r|r|r|r|r|r|r||r|r|r|r}
%ID & Multiplication & $m$ & $n$ & $k$ & \flops{} & \maxrowflops{} &$\|$C$\|$ & \maxrowsize & CF & CMRF &  Power8 & P100 & \begin{tabular}[c]{@{}r@{}}KNL\\ DDR\end{tabular} & \begin{tabular}[c]{@{}r@{}}KNL\\ CACHE\end{tabular} \\ 
%\hline
43 & hood & 220,542 & 220,542 & 220,542 & 562,028,138 & 3,871 & 34,242,180 & 231 & 0.13 & 0.16 & 15.28 & 25.34 & 9.44 & 12.45 \\ \hline
44 & Laplace\_284\_RxA & 2,774,624 & 22,906,304 & 22,906,304 & 582,550,744 & 350 & 206,478,136 & 114 & 0.72 & 0.73 & 2.13 & 4.58 & 0.87 & 2.05 \\ \hline
45 & Laplace\_284\_AxP & 22,906,304 & 22,906,304 & 2,774,624 & 582,550,744 & 36 & 206,478,136 & 15 & 0.93 & 1.00 & 3.42 & 6.76 & 2.40 & 2.89 \\ \hline
46 & af\_shell1 & 504,855 & 504,855 & 504,855 & 613,607,875 & 1,375 & 47,560,375 & 105 & 0.11 & 0.15 & 11.53 & 26.17 & 8.55 & 9.78 \\ \hline
47 & pwtk & 217,918 & 217,918 & 217,918 & 626,054,402 & 8,474 & 32,772,236 & 384 & 0.09 & 0.10 & 16.43 & 33.24 & 10.79 & 13.60 \\ \hline
48 & delaunay\_n24 & 16,777,216 & 16,777,216 & 16,777,216 & 633,914,372 & 280 & 347,322,258 & 218 & 0.47 & 0.69 & 3.57 & 6.72 & 2.23 & 3.11 \\ \hline
49 & Laplace\_284\_RxAP & 2,774,624 & 22,906,304 & 2,774,624 & 742,456,340 & 422 & 86,570,980 & 49 & 0.83 & 0.91 & 2.34 & 6.19 & 1.05 & 2.11 \\ \hline
50 & Laplace\_284\_RAxP & 2,774,624 & 22,906,304 & 2,774,624 & 742,456,340 & 404 & 86,570,980 & 49 & 0.93 & 0.97 & 2.84 & 4.76 & 1.02 & 2.11 \\ \hline
51 & Brick\_185\_RxA & 238,328 & 6,331,625 & 6,331,625 & 776,170,999 & 5,751 & 78,955,509 & 509 & 0.35 & 0.35 & 6.42 & 7.98 & 4.43 & 6.10 \\ \hline
52 & Brick\_185\_AxP & 6,331,625 & 6,331,625 & 238,328 & 776,170,999 & 215 & 78,955,509 & 27 & 0.61 & 0.83 & 6.94 & 11.70 & 4.69 & 5.66 \\ \hline
53 & nlpkkt80 & 1,062,400 & 1,062,400 & 1,062,400 & 790,384,704 & 784 & 154,663,144 & 152 & 0.38 & 0.48 & 7.64 & 12.43 & 6.10 & 7.17 \\ \hline
54 & BigStar\_4657\_AxP & 21,687,649 & 21,687,649 & 1,446,620 & 845,292,479 & 42 & 131,484,200 & 9 & 0.78 & 0.98 & 4.14 & 9.60 & 2.98 & 3.57 \\ \hline
55 & BigStar\_4657\_RxA & 1,446,620 & 21,687,649 & 21,687,649 & 845,292,479 & 585 & 131,484,200 & 91 & 0.40 & 0.42 & 3.82 & 7.71 & 2.15 & 4.08 \\ \hline
56 & eu-2005 & 862,664 & 862,664 & 862,664 & 849,268,919 & 153,454 & 284,177,131 & 16,051 & 0.28 & 0.10 & 5.48 & 5.44 & 2.06 & 2.78 \\ \hline
57 & NALU\_R3\_RxAP & 552,583 & 17,598,889 & 552,583 & 1,254,217,679 & 3,357 & 31,098,707 & 59 & 0.46 & 0.68 & 5.27 & 8.91 & 3.94 & 5.16 \\ \hline
58 & NALU\_R3\_RAxP & 552,583 & 17,598,889 & 552,583 & 1,254,217,679 & 3,219 & 31,098,707 & 59 & 0.67 & 0.77 & 4.75 & 4.78 & 3.40 & 4.56 \\ \hline
59 & Empire\_RxA & 8,800 & 2,160,000 & 2,160,000 & 1,286,511,829 & 155,460 & 25,410,400 & 3,010 & 0.15 & 0.17 & 8.16 & 8.66 & 5.55 & 6.56 \\ \hline
60 & Empire\_AxP & 2,160,000 & 2,160,000 & 8,800 & 1,286,511,829 & 999 & 25,410,400 & 27 & 0.55 & 0.28 & 8.36 & 8.85 & 6.46 & 7.42 \\ \hline
61 & Fault\_639 & 638,802 & 638,802 & 638,802 & 1,298,780,298 & 15,813 & 126,633,024 & 897 & 0.19 & 0.16 & 11.06 & 17.47 & 8.63 & 10.46 \\ \hline
62 & channel-500x100x100-b050 & 4,802,000 & 4,802,000 & 4,802,000 & 1,522,677,096 & 324 & 436,529,632 & 93 & 0.52 & 0.74 & 6.31 & 11.41 & 4.77 & 6.00 \\ \hline
63 & wb-edu & 9,845,725 & 9,845,725 & 9,845,725 & 1,559,579,990 & 281,616 & 630,077,764 & 14,427 & 0.21 & 0.12 & 4.25 & 7.07 & 1.31 & 1.40 \\ \hline
64 & Elasticity\_113\_AxP & 4,328,691 & 4,328,691 & 54,872 & 1,572,091,911 & 375 & 53,338,743 & 27 & 0.61 & 0.74 & 8.31 & 12.86 & 6.39 & 7.55 \\ \hline
65 & Elasticity\_113\_RxA & 54,872 & 4,328,691 & 4,328,691 & 1,572,091,911 & 30,375 & 53,338,743 & 1,029 & 0.14 & 0.14 & 9.43 & 12.64 & 7.23 & 8.93 \\ \hline
66 & in-2004 & 1,382,908 & 1,382,908 & 1,382,908 & 1,708,503,481 & 240,257 & 213,255,458 & 9,997 & 0.08 & 0.09 & 4.59 & 5.80 & 1.85 & 2.07 \\ \hline
67 & af\_shell10 & 1,508,065 & 1,508,065 & 1,508,065 & 1,840,916,875 & 1,225 & 142,742,975 & 95 & 0.11 & 0.16 & 12.10 & 27.42 & 9.99 & 11.49 \\ \hline
68 & cage15 & 5,154,859 & 5,154,859 & 5,154,859 & 1,885,387,372 & 1,904 & 929,023,247 & 859 & 0.83 & 0.79 & 4.57 & 5.72 & 2.78 & 4.33 \\ \hline
69 & dielFilterV2real & 1,157,456 & 1,157,456 & 1,157,456 & 2,337,362,192 & 5,464 & 325,027,200 & 668 & 0.30 & 0.36 & 8.76 & 13.36 & 7.12 & 8.64 \\ \hline
70 & wikipedia-20051105 & 1,634,989 & 1,634,989 & 1,634,989 & 2,373,873,670 & 274,992 & 1,725,264,272 & 134,091 & 0.82 & 0.19 & 3.33 &  & 0.99 & 1.16 \\ \hline
71 & ldoor & 952,203 & 952,203 & 952,203 & 2,408,881,377 & 4,165 & 145,422,935 & 259 & 0.12 & 0.16 & 12.22 & 24.74 & 10.70 & 12.08 \\ \hline
72 & NALU\_R3\_RxA & 552,583 & 17,598,889 & 17,598,889 & 2,456,399,623 & 6,777 & 243,321,996 & 613 & 0.28 & 0.32 & 6.37 & 8.20 & 4.44 & 6.49 \\ \hline
73 & NALU\_R3\_AxP & 17,598,889 & 17,598,889 & 552,583 & 2,456,399,623 & 183 & 243,321,996 & 22 & 0.67 & 0.90 & 5.38 & 11.50 & 4.37 & 4.93 \\ \hline
74 & Serena & 1,391,349 & 1,391,349 & 1,391,349 & 3,111,966,351 & 11,556 & 315,805,689 & 1,236 & 0.18 & 0.18 & 11.33 & 17.13 & 10.07 & 11.26 \\ \hline
75 & coPapersDBLP & 540,486 & 540,486 & 540,486 & 4,091,407,036 & 737,309 & 480,122,442 & 35,874 & 0.17 & 0.03 & 10.18 & 7.89 & 5.20 & 7.81 \\ \hline
76 & flickr & 820,878 & 820,878 & 820,878 & 4,318,945,024 & 2,907,529 & 1,112,536,788 & 202,990 & 0.57 & 0.03 & 4.45 & 1.45 & 1.28 & 2.17 \\ \hline
77 & RM07R & 381,689 & 381,689 & 381,689 & 5,272,142,064 & 60,375 & 193,345,783 & 1,475 & 0.15 & 0.15 & 11.81 & 20.33 & 10.08 & 12.15 \\ \hline
78 & Bump\_2911\_dig & 2,911,419 & 2,911,419 & 2,911,419 & 5,745,156,927 & 9,370 & 560,173,611 & 738 & 0.17 & 0.17 & 11.29 & 19.54 & 10.49 & 11.39 \\ \hline
79 & kron\_g500-logn16 & 65,536 & 65,536 & 65,536 & 6,768,428,563 & 4,368,565 & 972,785,311 & 54,856 & 0.50 & 0.03 & 8.44 & 2.49 & 2.66 & 3.63 \\ \hline
80 & coPapersCiteseer & 434,102 & 434,102 & 434,102 & 6,822,448,658 & 786,040 & 264,584,716 & 12,882 & 0.10 & 0.03 & 8.81 & 19.43 & 5.59 & 7.50 \\ \hline
81 & audikw\_1 & 943,695 & 943,695 & 943,695 & 8,089,734,897 & 32,985 & 662,878,935 & 1,689 & 0.14 & 0.14 & 11.71 & 19.70 & 9.92 & 11.80 \\ \hline
82 & dielFilterV3real & 1,102,824 & 1,102,824 & 1,102,824 & 8,705,461,058 & 26,163 & 688,649,400 & 1,671 & 0.23 & 0.26 & 10.47 & 17.04 & 8.60 & 10.83 \\ \hline
83 & HV15R & 2,017,169 & 2,017,169 & 2,017,169 & 42,201,218,799 & 132,295 & 1,768,066,720 & 1,900 & 0.13 & 0.14 & 13.00 &  & 10.25 & 8.22 \\ \hline
 &  &  &  &  &  &  &  &  GEOMEAN: & 0.38 & 0.30 & 5.23 & 5.98 & 2.86 & 3.90\end{tabular}
}
\end{table}

%We compare {\sc kkmem} with \spgemm{} implementations in 
%cuSPARSE~\cite{cusparseurl}, CUSP~\cite{dalton2015optimizing}, 
%bhSPARSE~\cite{liu2014efficient}, and ViennaCL ~\cite{Rupp:ViennaCL},
%and AmgX~\cite{demouth2012sparse} on \gpu{}s (Section~\ref{sec:gpuexp}).
%Intel Math Kernel Library (\mkl{}) and ViennaCL are used for comparisons
%on \cpu{}s and \knl{}s.
%Within these experiments, we also study the practical use case of reusing the 
%symbolic structure.

\myspace{-3ex} 
\subsection{Experiments on Power8 \cpu{}s}
\label{sec:cpuexp}

\myspace{-1.25ex}

\begin{figure}
\begin{center}

\subfloat[BigStar $A\times P$ $k=1.5M$]
{\includegraphics[width=0.33\columnwidth]{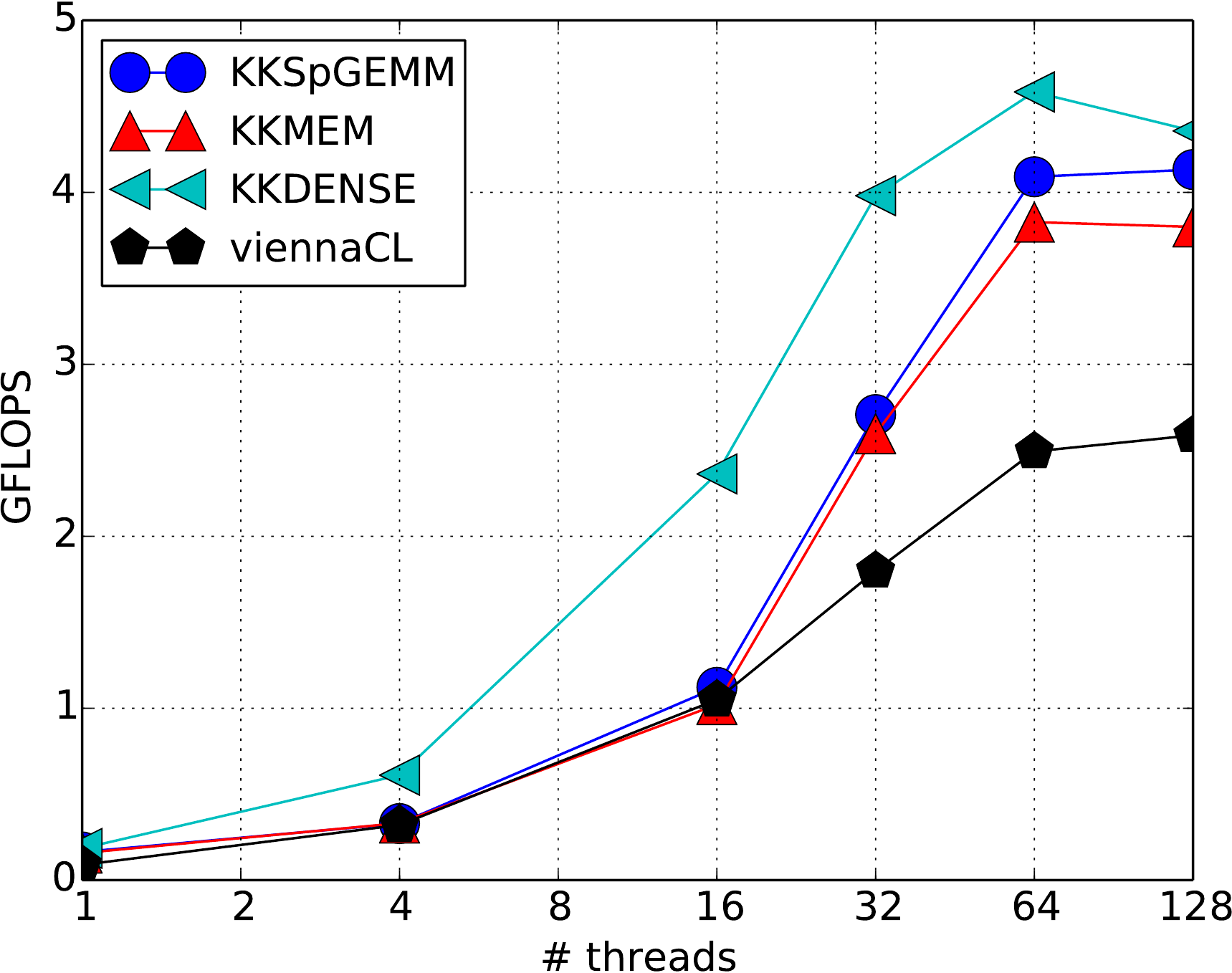}\label{fig:p8bap}}
\subfloat[BigStar $R\times A$ $k =21.5M$]
{\includegraphics[width=0.33\columnwidth]{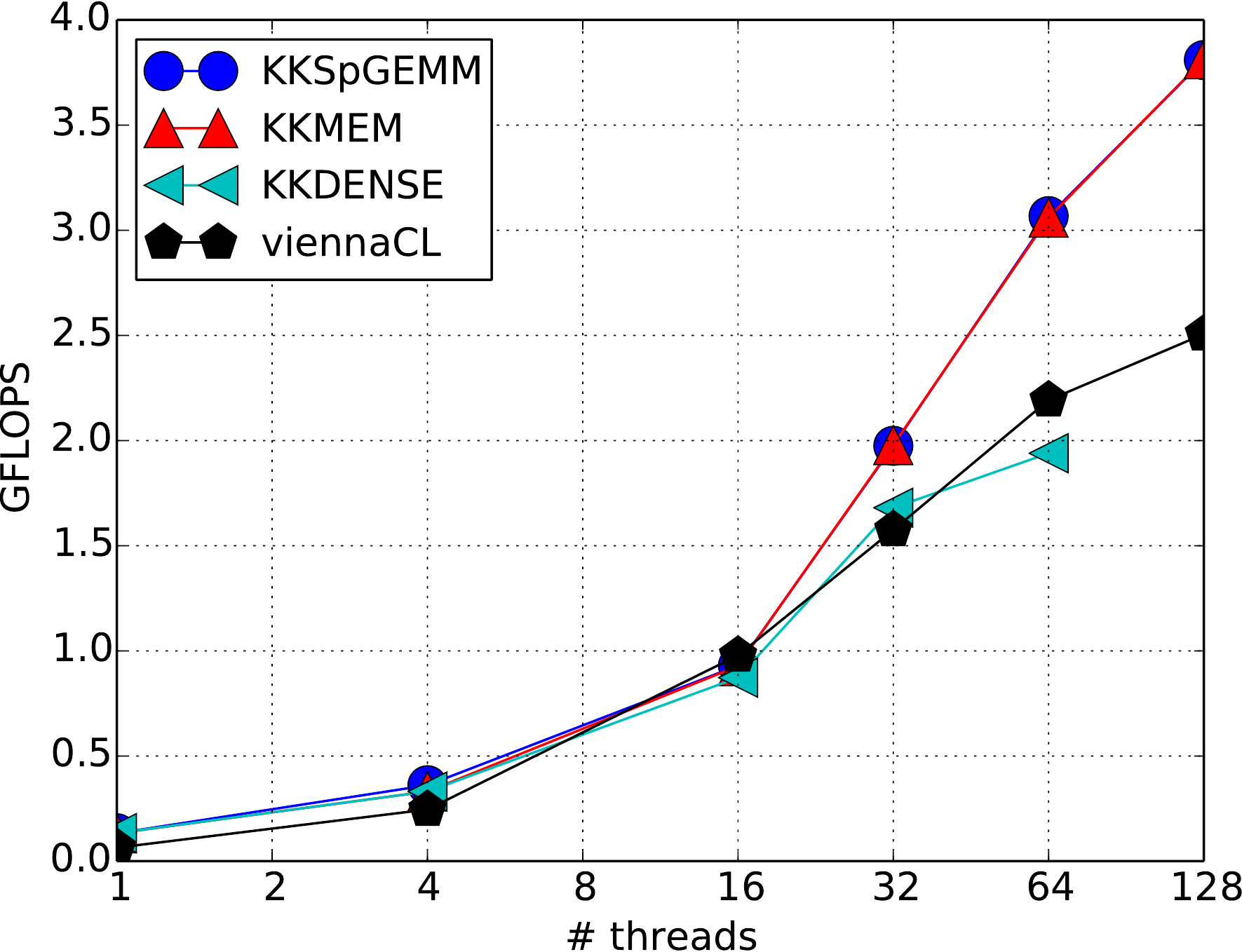}\label{fig:p8bra}}
\subfloat[europe $k = 50.9M$]
{\includegraphics[width=0.33\columnwidth]{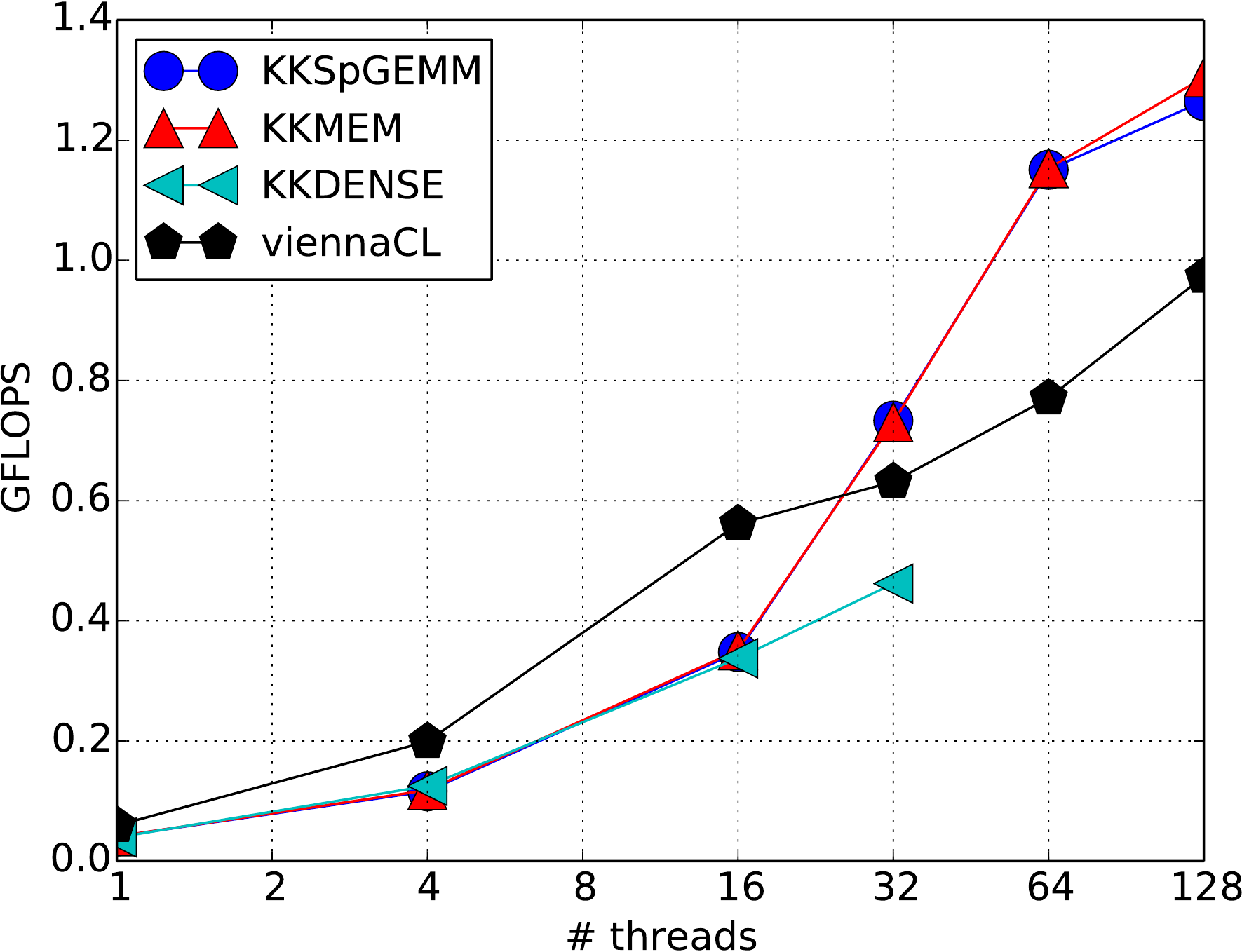}\label{fig:p8eu}}

\myspace{-3ex} 
\subfloat[kron16 $k = 65K$]
{\includegraphics[width=0.33\columnwidth]{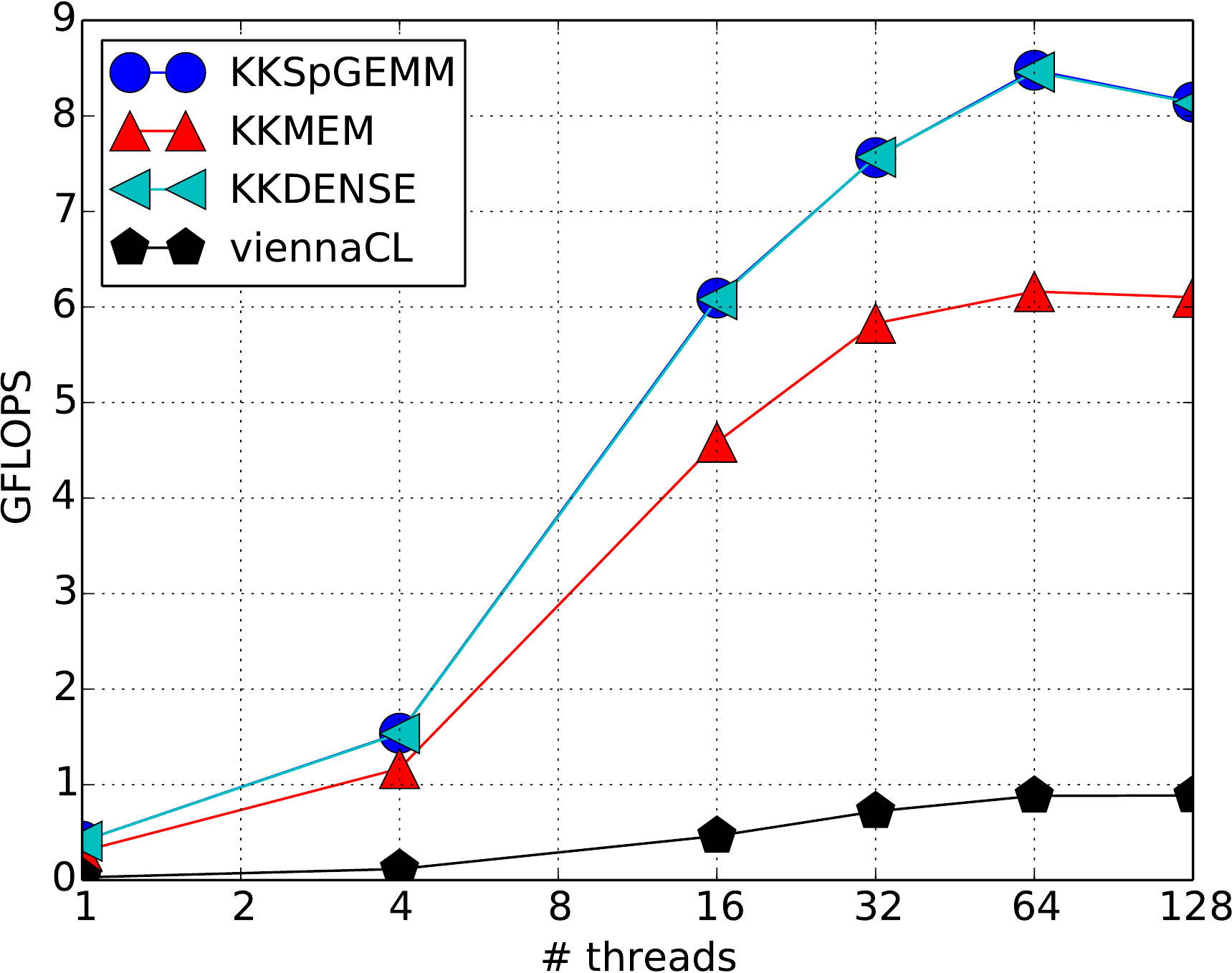}\label{fig:p8kr}}
\subfloat[coPap.Cite. $k=434K$]
{\includegraphics[width=0.33\columnwidth]{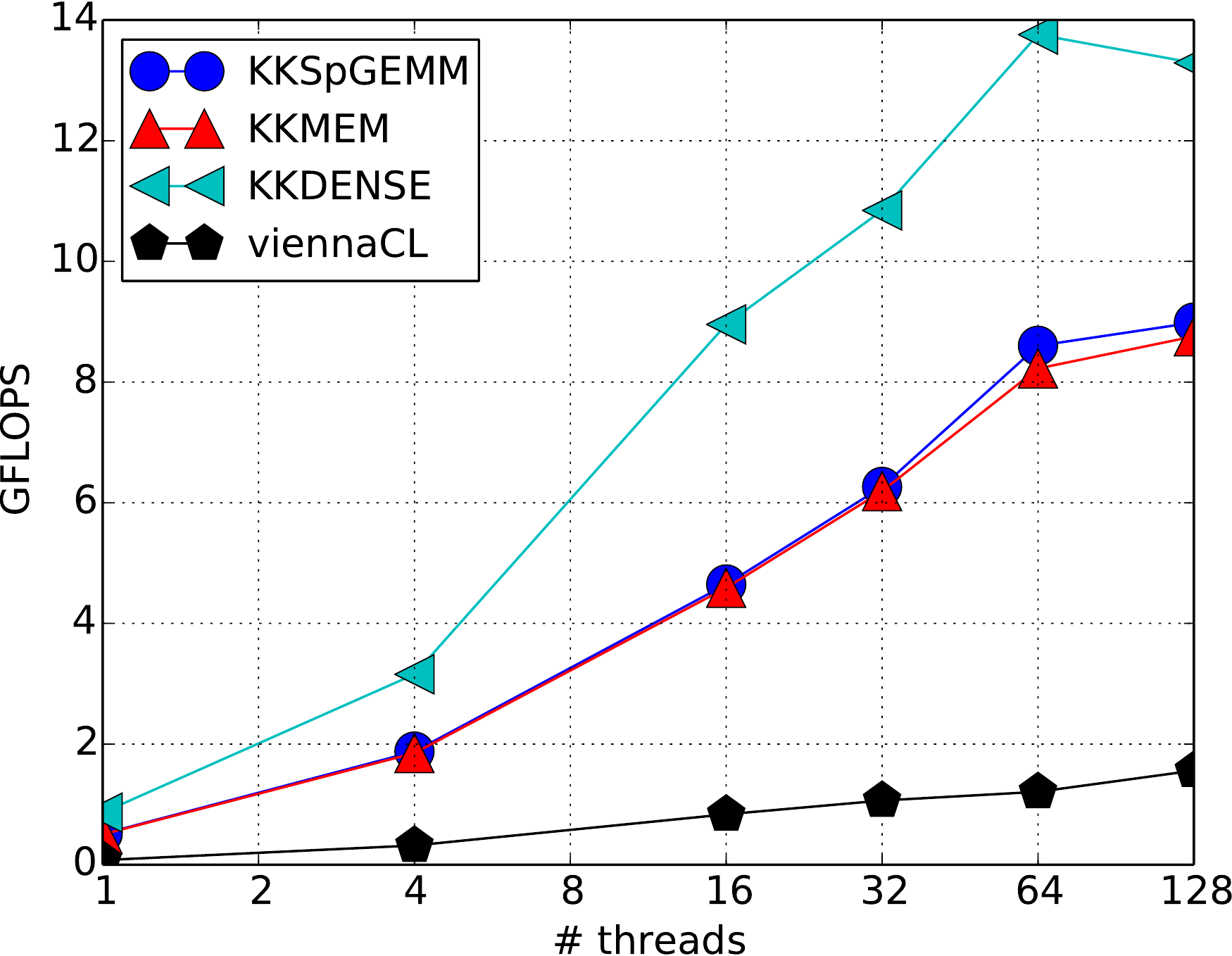}\label{fig:co}}
\subfloat[flickr $k = 820K$]
{\includegraphics[width=0.33\columnwidth]{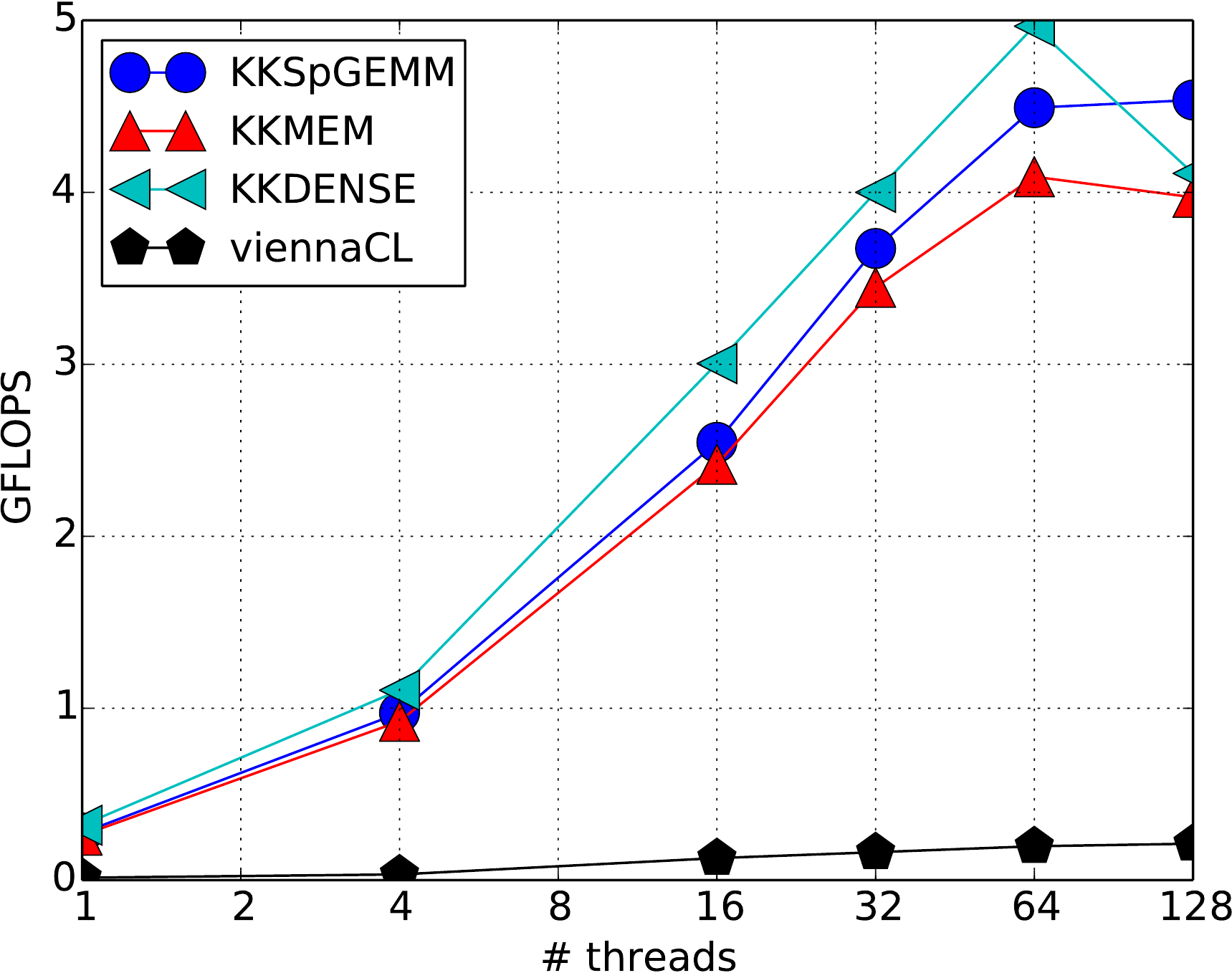}\label{fig:p8f}}
\end{center}
\myspace{-2ex} 
\caption{ \verysmallfont
Strong scaling GFLOPS/sec on Power8 \cpu{}s.  }
\label{fig:power8scale}
\end{figure}

We compare our methods ({\sc kkspgemm}, {\sc kkmem}, {\sc kkdense}), 
against ViennaCL (OpenMP) on Power8 \cpu{}s. Figure~\ref{fig:power8scale} gives
strong scaling GFLOPS/sec for the four methods on different multiplications with different 
characteristics. 

The first two multiplications (a and b) are from a multigrid problem. As $k$ gets larger, 
{\sc kkdense} suffers from low spatial locality, and it is outperformed by {\sc kkmem}.
{\sc kkdense}'s memory allocation for its accumulators fails for some cases.
Although, they should fit into memory, we suspect that allocation of
such large chunks is causing these failures. {\sc kkdense} achieves
better performance for matrices with smaller $k$. Among them,
kron not only has the smallest $k$, but also has a \maxrowsize{} that is $83\%$ of $k$. 
The sparse accumulators use a similar amount of memory as {\sc kkdense}, but still acrue the overhead for hash operations.
Our meta method chooses {\sc kkdense} for kron's numeric and the symbolic phase. 
It executes {\sc kkdense} only for the symbolic phase of BigStar $A\times P$, coPapersCiteseer, and flickr, as 
the compression reduces their $k$ by $32\times$. 
{\sc kkdense} achieves better performance than {\sc kkspgemm} in $3$ instances. 
These suggest that the simple architecture agnostic heuristic for choosing the optimal algorithm,
is leaving room for improvement.
%That said, the 
The current heuristic is erring on the side of reduced memory consumption, which
in real applications may be desirable. 
%With a better knowledge of the underlying architecture and the structure of the problem,
%a better cut-off parameter can be specified. 
%This requires more architecture specific 
%experimentation, and we leave it for future work. 
%In this work, 

\begin{figure}
\begin{center}

\subfloat[Power8 NoReuse]
{\includegraphics[width=0.50\columnwidth]{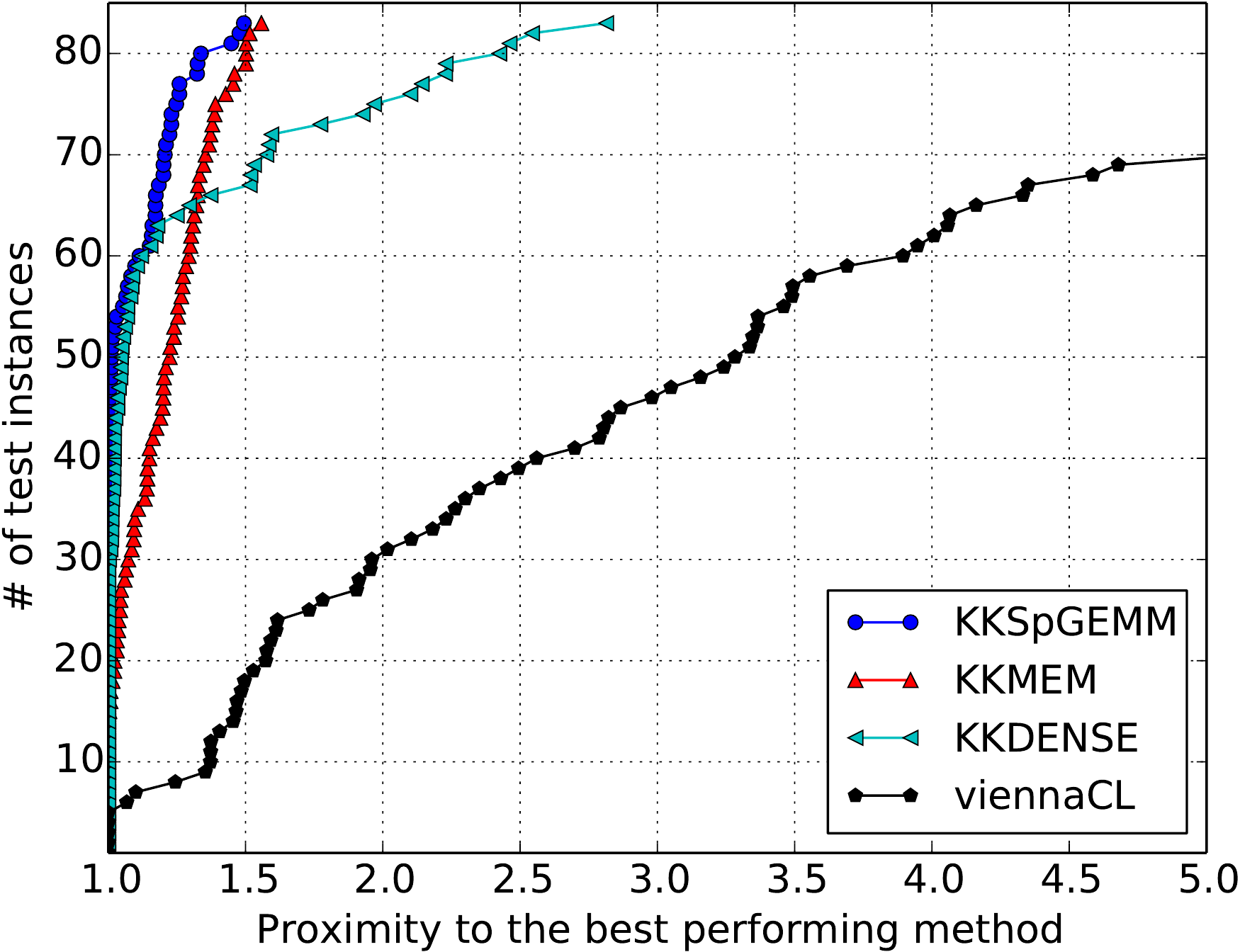}\label{fig:power8pp}}
\subfloat[\knl{} {\sc ddr} NoReuse ]
{\includegraphics[width=0.50\columnwidth]{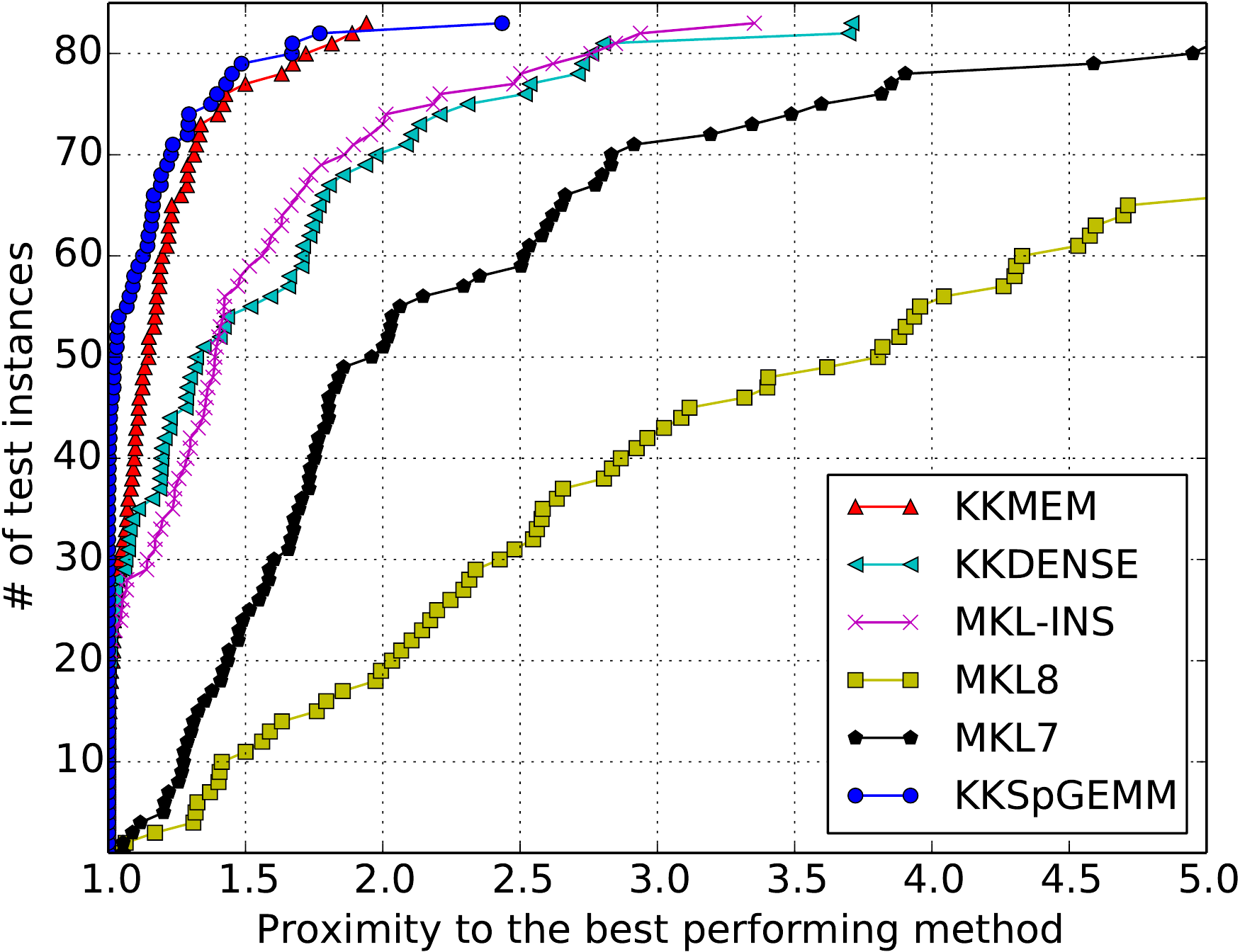}\label{fig:knlddroveral}}

\subfloat[\knl{} CM NoReuse]
{\includegraphics[width=0.50\columnwidth]{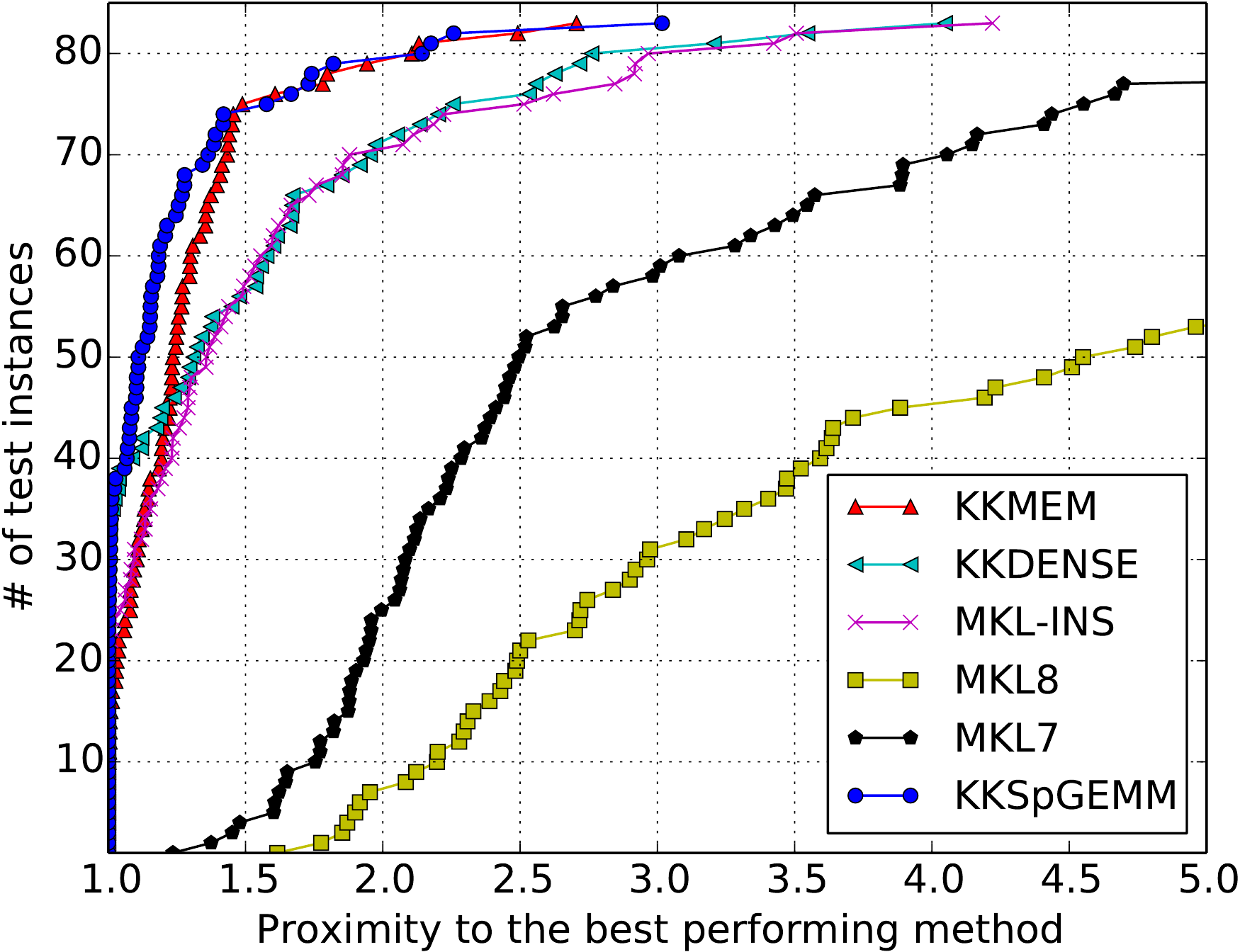}\label{fig:knlcacheoveral}}
\subfloat[\knl{} CM Reuse]
{\includegraphics[width=0.50\columnwidth]{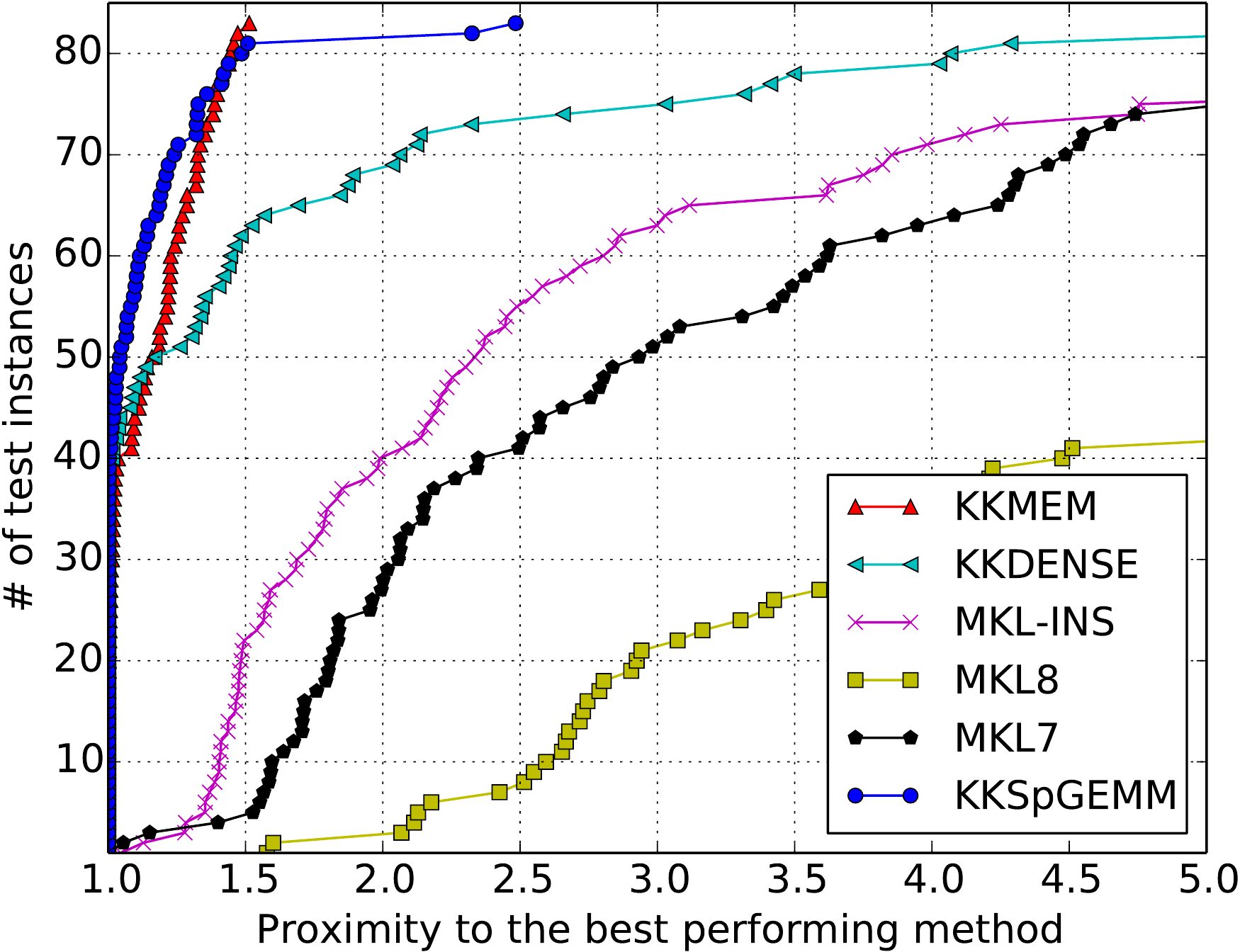}\label{fig:knlcachenumeric}}

\subfloat[P100 NoReuse]
{\includegraphics[width=0.50\columnwidth]{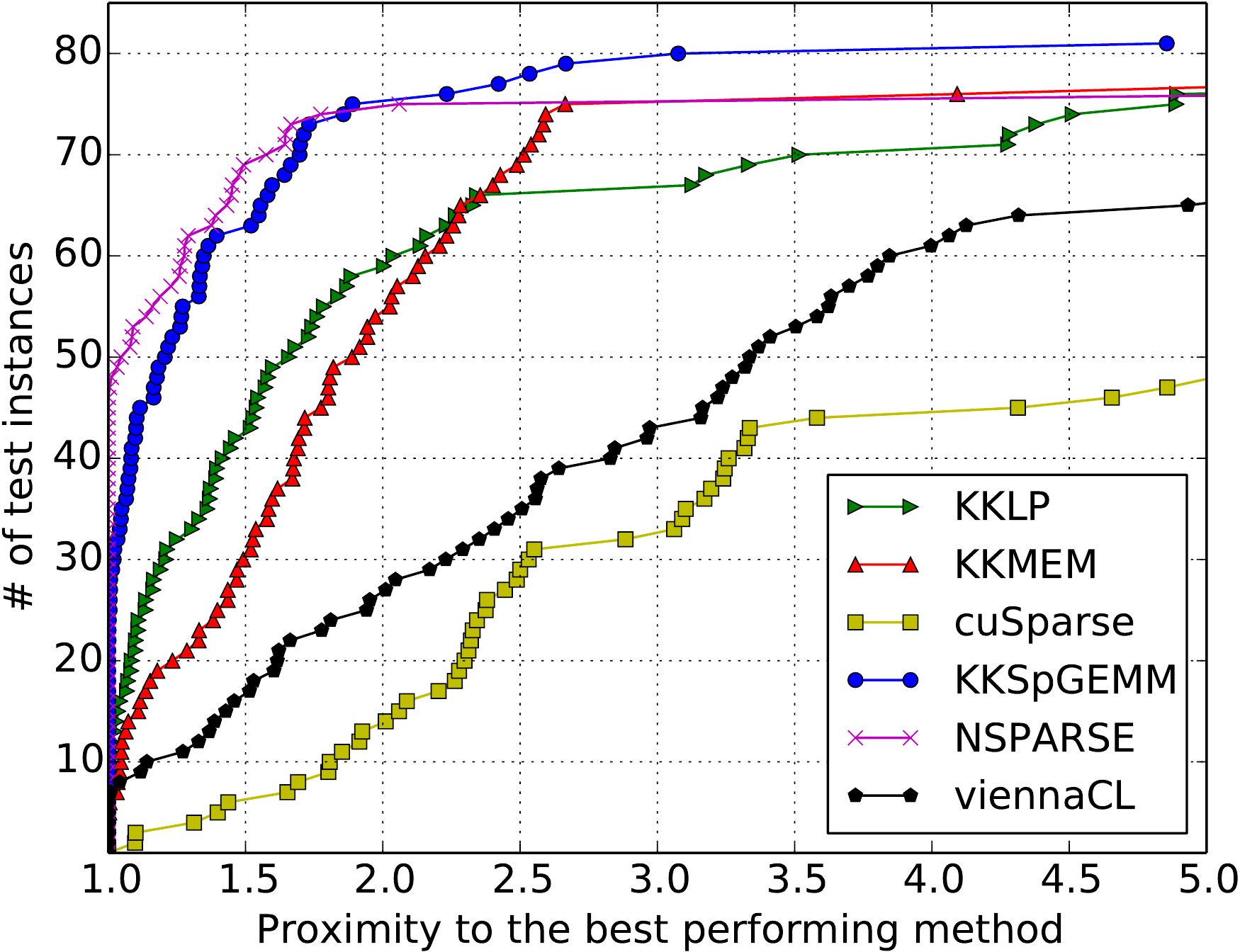}\label{fig:p100ppnr}}
\subfloat[P100 Reuse]
{\includegraphics[width=0.50\columnwidth]{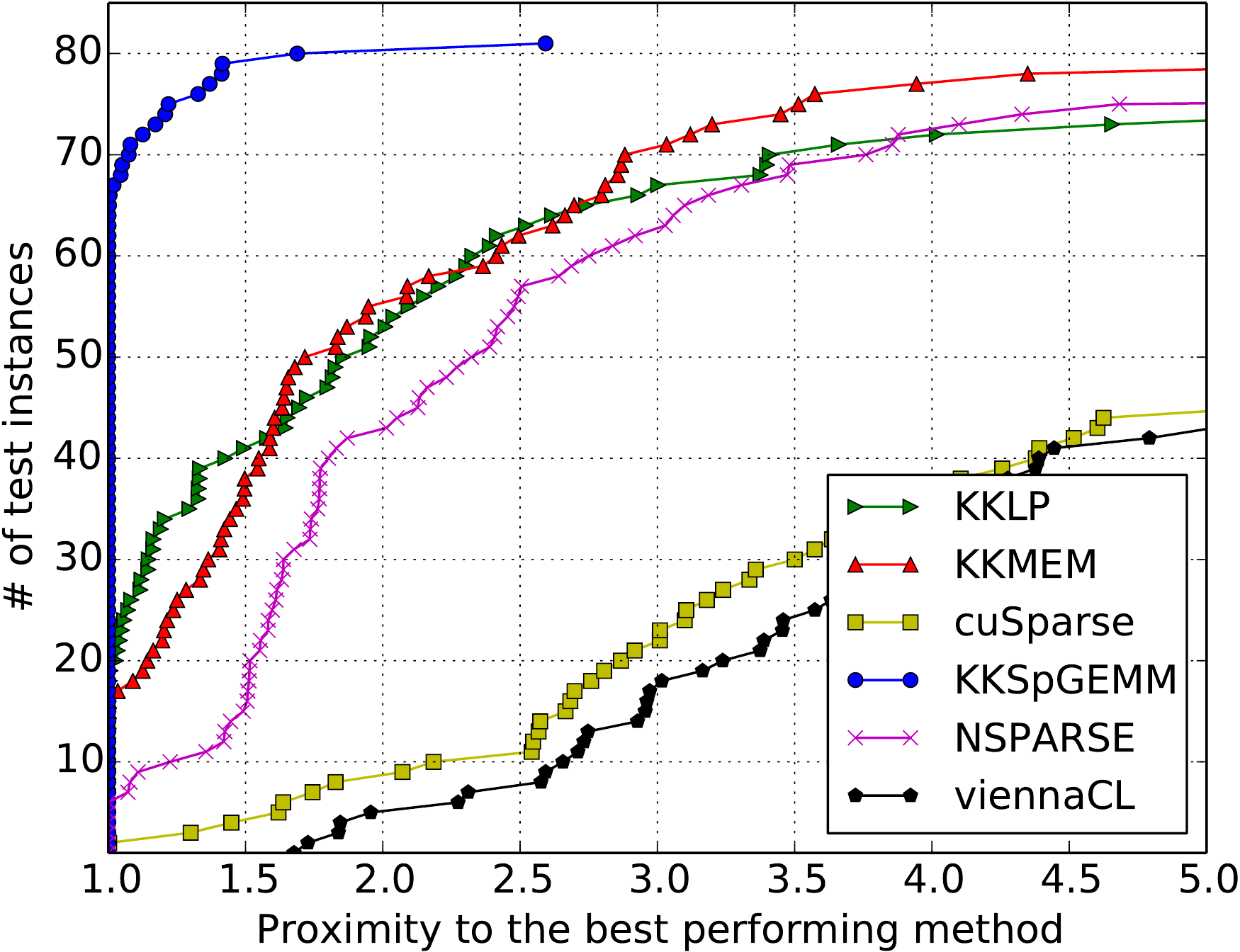}\label{fig:p100ppr}}
\end{center}
\caption{ \verysmallfont
Performance profiles on Power8, \knl{}, and P100 \gpu{}s. 
%Both symbolic and numeric is executed for NoReuse. 
%Reuse case executes only numeric phase of the methods (for those that are two-phase), and 
%reuses the previous symbolic computations. 
Experiments on \gpu{}s include 
$81$ multiplications, as result $C$ does not fit into memory.
}
\label{fig:knlpp}
\end{figure}

Figure~\ref{fig:power8pp} lists the performance profiles of the algorithms on Power8.
%and Reuse cases. 
For a given $x$, the $y$ value indicates the number of problem cases, for which a method is less
than $x$ times slower than the best result achieved with any method for each individual problem. 
The max value of $y$ at $x=1$ is the number of problem cases for which a method achieved the best performance.
The $x$ value for which $y=83$ is the largest slowdown a method showed over any problem, compared with the
best observed performance for that problem over all methods. 
As seen in the figure, for about $50$ problems {\sc kkspgemm} achieves
the best performance (or at most $0.5\%$ slower than the best KK variant). 
The performance of viennaCL is mostly lower than achieved by KK variants. 
%This performance 
%difference increases for Reuse case. Although viennaCL is a two-phase algorithm,
%the user interface do not allow support this feature. This highlights the importance of 
%the interface design for scientific computing problems. 
While our methods do not make any assumption on whether the input matrices have sorted columns,
all test problems have sorted columns to be able to run the different methods throughout our 
experiments. For example, viennaCL requires sorted inputs, and returns sorted output.
If the calling application does not store sorted matrices, pre-processing is required 
to use viennaCL. Similarly, if the result of \spgemm{} must be sorted, 
post-processing is required for our methods. For iterative multiplications in multigrid, 
the output of a multiplication ($AP=A \times P$) becomes the input of the next one ($R\times AP$).
As long as methods make consistent assumptions for their input and outputs, this pre-/post-processing
can be skipped.

%\begin{figure}
%\begin{center}
%

%\caption{ \verysmallfont
%Performance profile on P100 \gpu{}s for $81$ multiplications. $C$ does not fit into memory for two multiplications.}
%\label{fig:p100pp}
%\end{figure}

%\begin{figure}
%\begin{center}
%
%\subfloat[NoReuse]
%{\includegraphics[width=0.50\columnwidth]{experiment/white/openmp/perfprofchart/overalperf_POWER8}\label{fig:power8ppnr}}
%\subfloat[Reuse]
%{\includegraphics[width=0.50\columnwidth]{experiment/white/openmp/perfprofchart/numericperf_POWER8}\label{fig:power8ppr}}
%\end{center}
%\caption{ \verysmallfont
%Performance profile for all multiplications used in this paper on Power8. 
%}
%\label{fig:power8pp}
%\end{figure}

\myspace{-3ex}
\subsection{Experiments on \knl{}s}
\label{sec:knlexp}
\myspace{-1.25ex}

The experiments on \knl{}s compare our methods against two methods 
provided by the Intel Math Kernel Library (\mkl{}) using two memory modes. The first uses
the high bandwidth memory ({\sc mcdram}) of \knl{}s as a cache (CM), while the second 
runs in flat memory mode using only {\sc ddr}.
{\tt mkl\_sparse\_spmm} in \mkl{}'s inspector-executor is referred to as {\sc mkl-ins}, 
and  the {\tt mkl\_dcsrmultcsr} is referred to as {\sc mkl7} and {\sc mkl8}. 
{\tt mkl\_dcsrmultcsr} requires sorted inputs, without necessarily returning sorted 
outputs. Output sorting may be skipped for $A\times A$ (e.g., graph analytic problems);
however, it becomes an issue for multigrid. The results report
both {\sc mkl7} (without output sorting) and {\sc mkl8} (with output sorting).
%, in which{\sc mkl7} skips output sorting, while {\sc mkl8} performs this output sorting.
\mkl{}'s expected performance is the performance of {\sc mkl7} for $A\times A$ and {\sc mkl8} for 
multigrid multiplications.
%the performance of {\sc mkl7} and {\sc mkl8}  
%A general performance between the performance of these two can be expected from {\tt mkl\_dcsrmultcsr}.
%Our multiplications include both multigrid and $A\times A$ multiplications. A performance between 
%these two methods can be expected from {\tt mkl\_dcsrmultcsr}.
%Both \mkl{} routines are {up to \emph{$2-3\times$ slower}} for the first call 
%than the following calls in any run. %Even though an application
%using \mkl{} will observe this difference, we exclude the first run
%for these routines. Thus, {\emph{the comparisons against \mkl{} are conservative}}.

\begin{figure}
\begin{center}
\subfloat[coPapersCiteseer {\sc ddr}]
{\includegraphics[width=0.33\columnwidth]{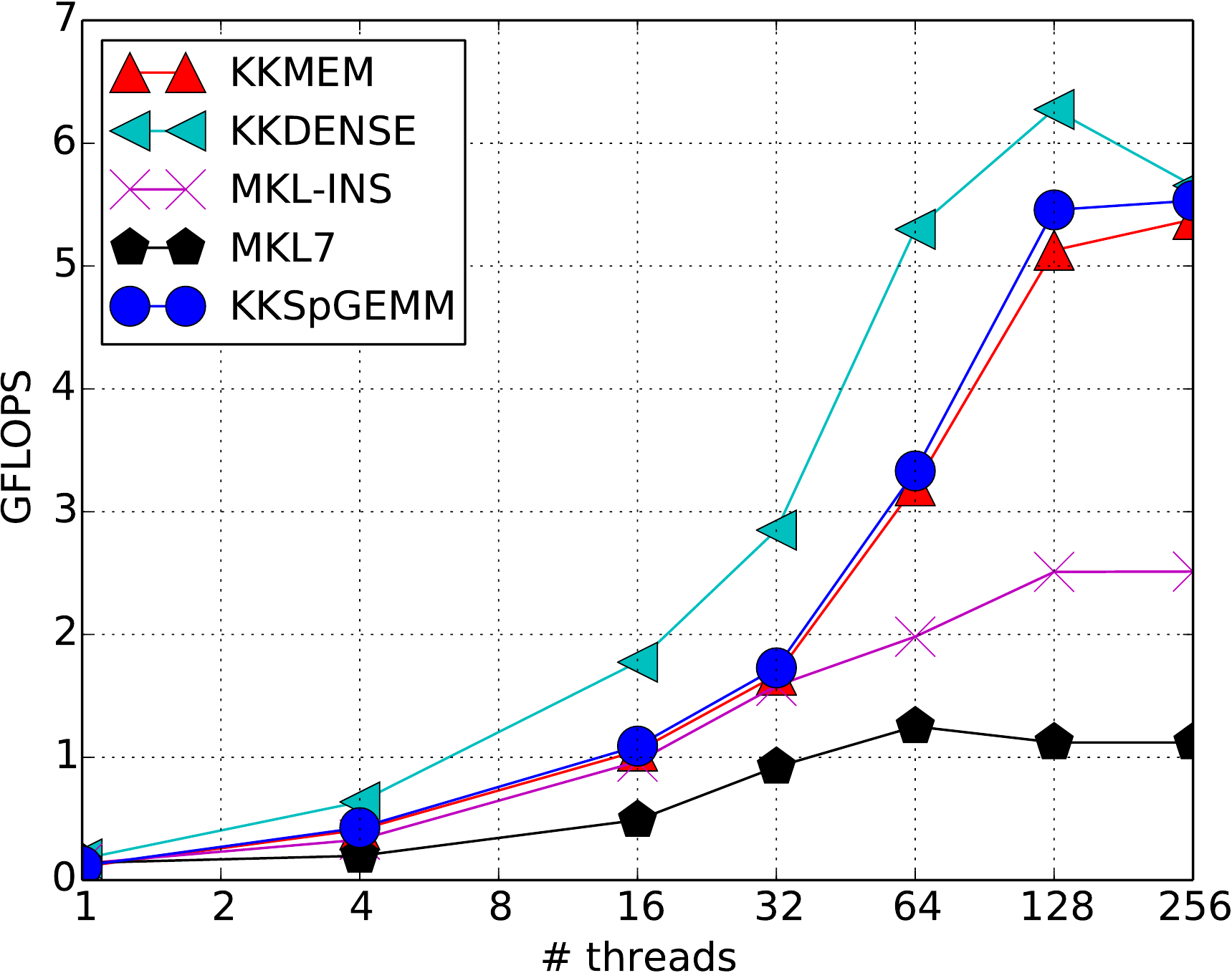}\label{fig:knlddrcpc}}
\subfloat[BigStar $A\times P$ {\sc ddr}]
{\includegraphics[width=0.33\columnwidth]{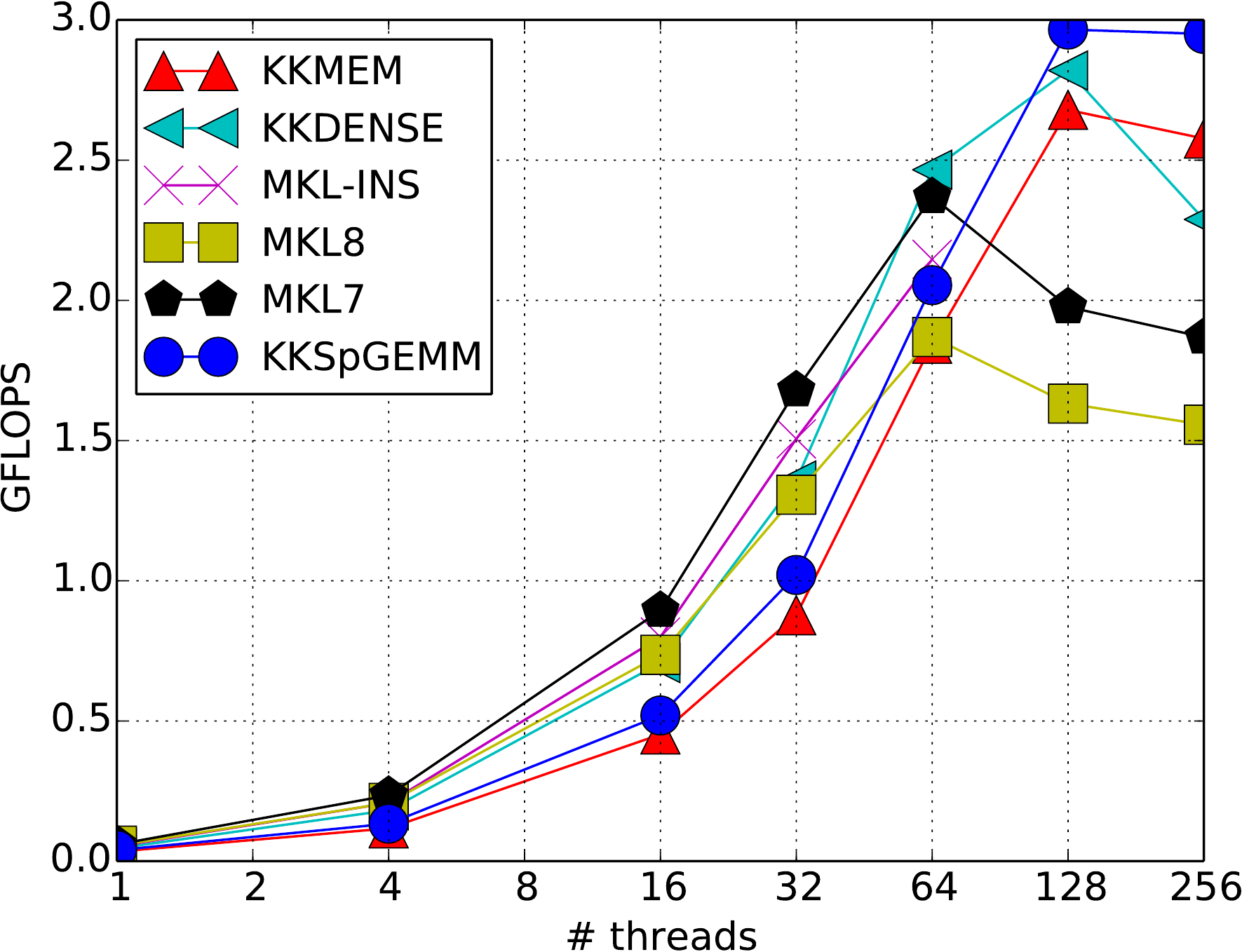}\label{fig:knlddrbap}}
\subfloat[BigStar $R\times A$ {\sc ddr}]
{\includegraphics[width=0.33\columnwidth]{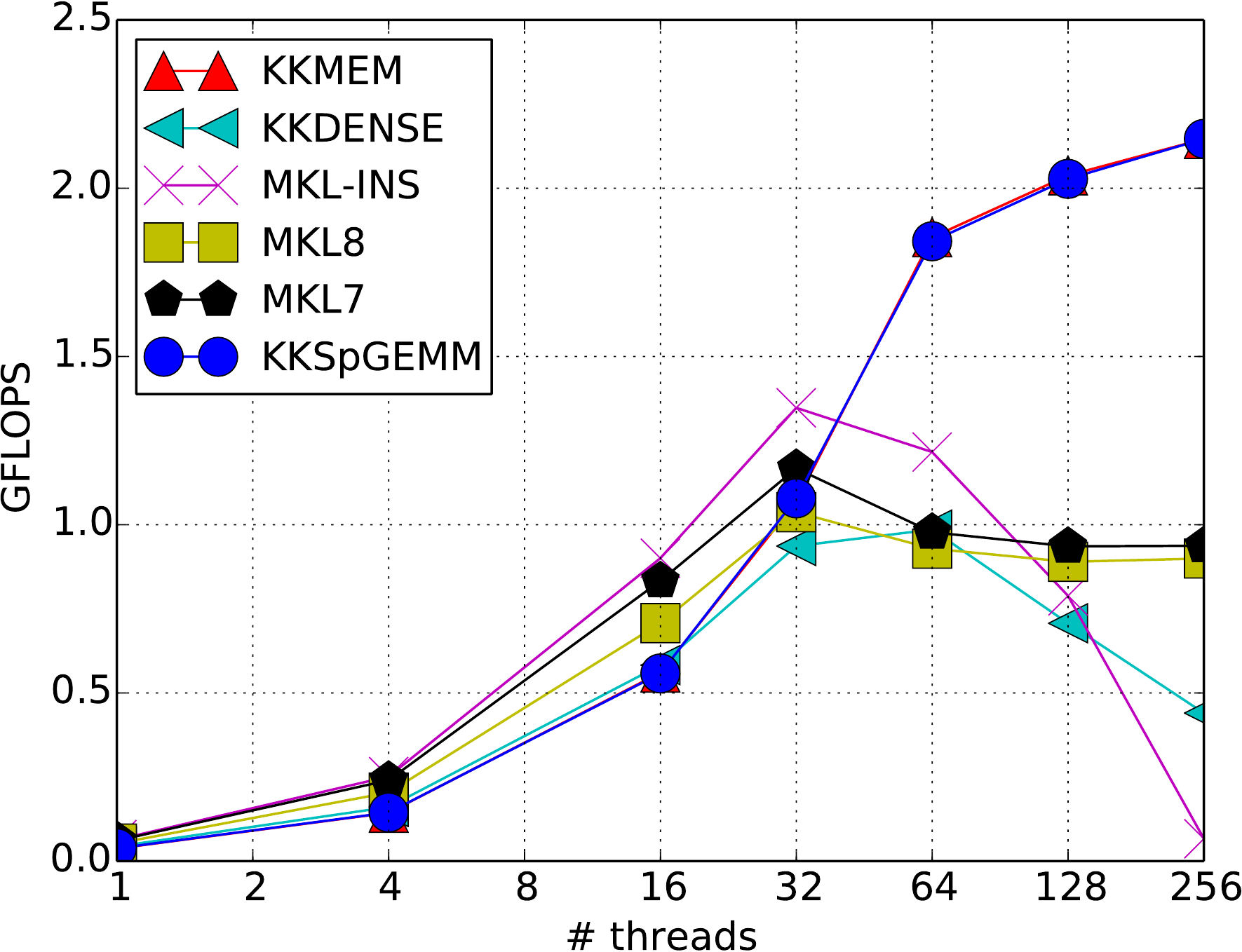}\label{fig:knlddrbra}}

\myspace{-3ex} 
\subfloat[coPapersCiteseer CM]
{\includegraphics[width=0.33\columnwidth]{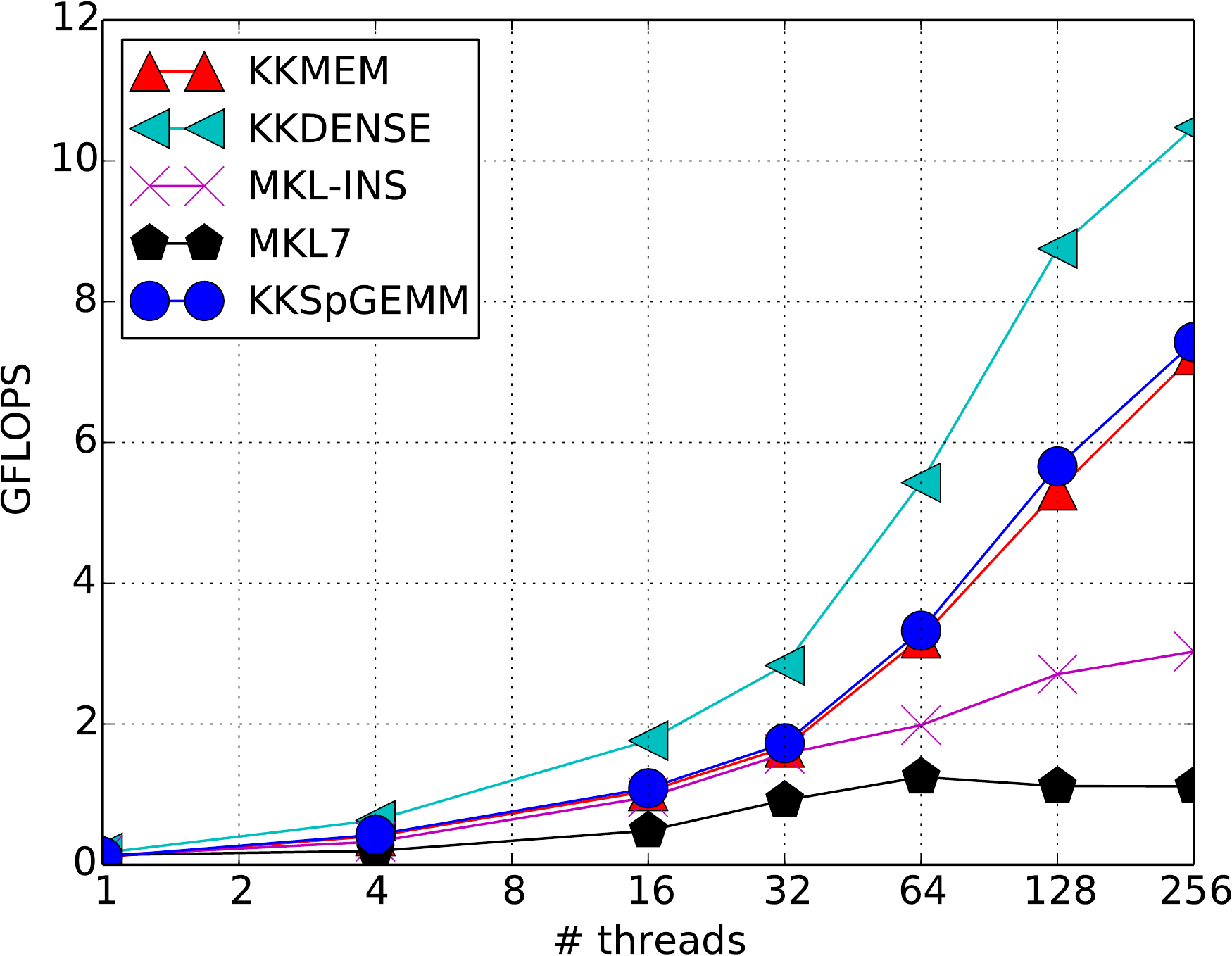}\label{fig:knldeltacpc}}
\subfloat[BigStar $A\times P$ CM]
{\includegraphics[width=0.33\columnwidth]{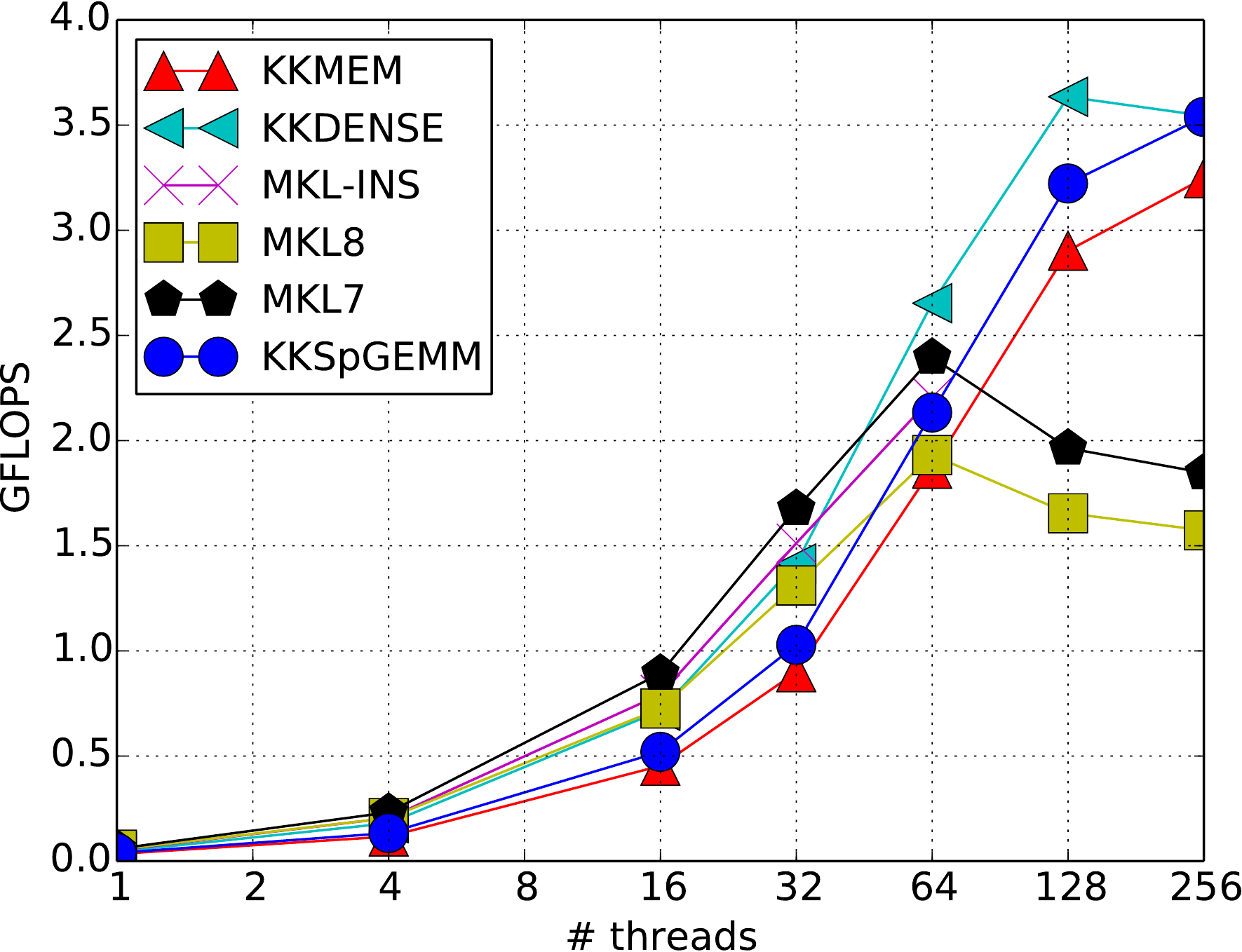}\label{fig:knldeltabap}}
\subfloat[BigStar $R\times A$ CM]
{\includegraphics[width=0.33\columnwidth]{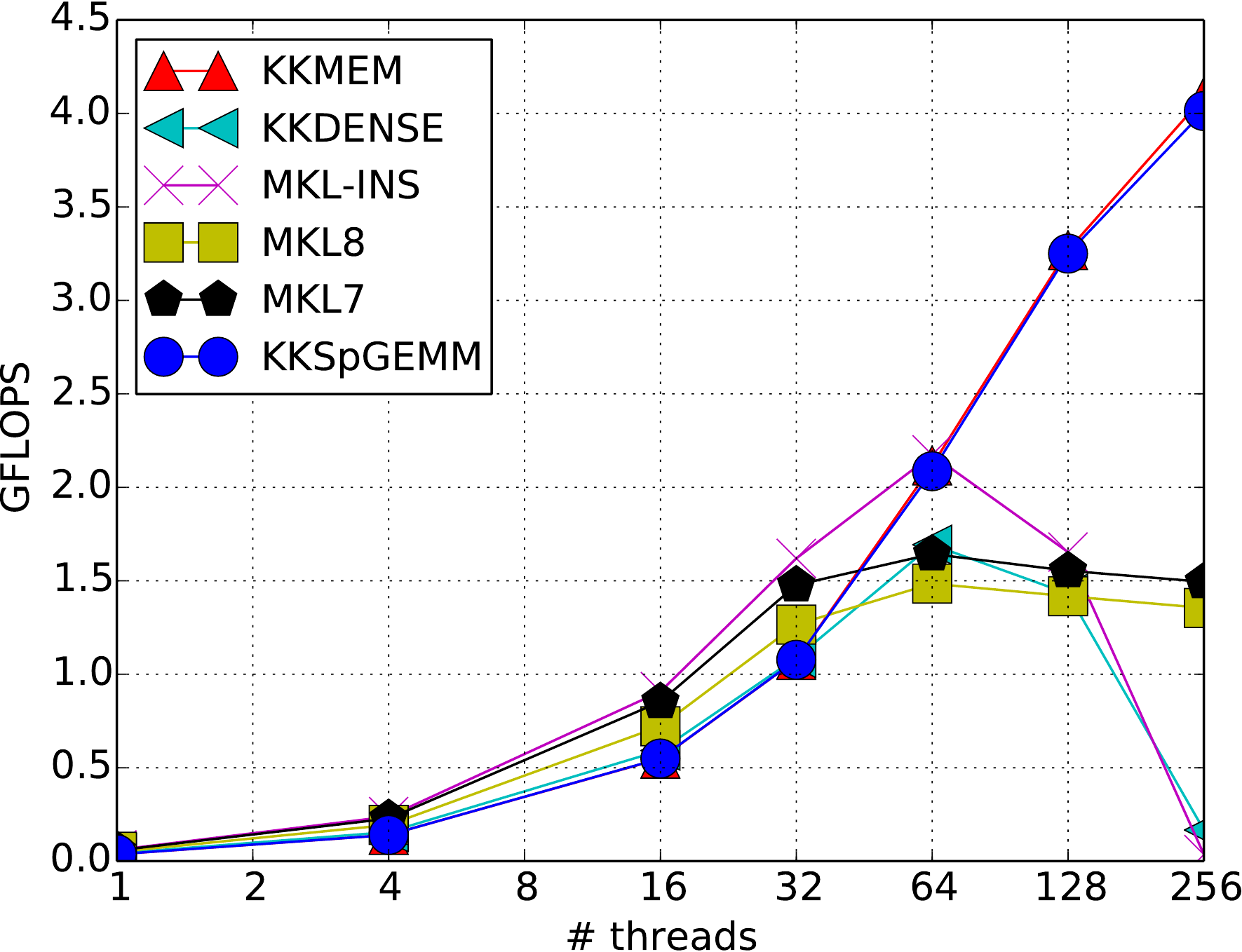}\label{fig:knldeltabra}}
\myspace{-3ex} 
\end{center}
\caption{ \verysmallfont
Strong scaling GLOPS/sec on \knl{}s. Top and bottom figures are for flat {\sc ddr} and CM, respectively.
{\sc mkl8} does not complete in the given allocation time for coPapersCiteseer.}
\label{fig:knlscale}
\end{figure}

Figure~\ref{fig:knlscale} shows strong scaling GFLOPS/sec of six methods 
on three multiplications for CM and {\sc ddr} with number of threads. Since we
use maximally 64 cores, 128 and 256 threads use 2 or 4 hyperthreads per cores
respectively.
Memory accesses on {\sc ddr} are 
relatively more expensive than CM; therefore methods using sparse accumulators 
tend to achieve better scaling and performance. {\sc kkdense} is usually 
outperformed by {\sc kkmem} on {\sc ddr} except on coPapersCiteseer. CM provides 
more bandwidth which boosts the performance of all methods. 
%{\sc mcdram} provides not only more bandwidth, but also 
%more parallelism for the memory controllers compared to {\sc ddr}. 
%When neither the bandwidth nor the memory controllers are saturated,
When the bandwidth is not saturated, 
methods have similar performances on {\sc ddr} and {\sc mcdram}, 
%the performance of methods is expected to be similar on {\sc ddr} and {\sc mcdram}, 
which is observed up to $32$ threads.
%suggesting that cores do not saturate the available bandwidth. 
%{\sc mcdram} clearly helps in this case. 
%Even when the algorithm is not 
%bandwidth bounded on {\sc ddr}, it may saturate memory controllers, 
%and may suffer from latency due to the memory request queues. {\sc mcdram} 
%can help here with more parallelism in the memory request queues.
CM improves the performance of methods which stress memory accesses more, 
e.g. {\sc kkdense}. In general, {\em methods favoring memory accesses over hash computations
are more likely to benefit from CM than those that already have localized memory accesses.}
%{\sc kkspgemm} uses the same cut-off value for both memory modes. 
{\sc kkspgemm} mostly achieves the best performance except for coPapersCiteseer. The higher memory bandwidth 
of CM allows the use of dense accumulators for larger $k$. 
$k$ is still too large to benefit from CM for $R\times A$. {\sc mkl} methods
achieve better performance on lower thread counts, but they do not scale 
with hyperthreads. {\sc mkl-ins} has the best performance among {\sc mkl} methods.

It is worthwhile to note that these thread scaling experiments conflate two performance critical 
issues: thread-scalability of an algorithm, and the amount of memory bandwidth and load/store slots
available to each thread. The latter issue would still afflict performance if these methods are
used as part of an MPI application, where for example 8 MPI ranks each use 32 threads on KNL. 
In such a usecase we would expect the relative performance of the methods to be closer to 
the $256$ thread case than the $32$ thread case in our experiments. 

Figure~\ref{fig:knlpp} shows performance profiles for NoReuse for {\sc ddr}, and both NoReuse and Reuse for CM.
The experiments on {\sc ddr} demonstrate the strength of a thread-scalable {\sc kkmem} 
algorithm. It outperforms {\sc kkdense} for larger datasets. Overall, {\sc kkspgemm} obtains 
the best performance, taking advantage of {\sc kkmem} and {\sc kkdense } for large and 
small datasets, respectively. {\sc kkdense} significantly improves its performance on CM w.r.t. 
{\sc ddr}. %The that cut-off parameter of {\sc kkspgemm} for $k$ can be pushed further for CM.
Among mkl methods, {\sc mkl-ins} achieves the best performance. However, it is a 1-phase method.
It cannot exploit structural reuse, and its performance drops for the Reuse case.

\myspace{-3ex}
\subsection{Experiments on GPUs}
\label{sec:gpuexp}
\myspace{-1.25ex}

%\begin{figure}
%\begin{center}
%
%\subfloat[NoReuse]
%{\includegraphics[width=0.50\columnwidth]{experiment/white/cuda/perfprofchart/overalperf_P100}\label{fig:p100ppnr}}
%\subfloat[Reuse]
%{\includegraphics[width=0.50\columnwidth]{experiment/white/cuda/perfprofchart/numericperf_P100}\label{fig:p100ppr}}
%\end{center}
%\caption{ \verysmallfont
%Performance profile on P100 \gpu{}s for $81$ multiplications. $C$ does not fit into memory for two multiplications.}
%\label{fig:p100pp}
%\end{figure}

We evaluate the performance of our methods against Nsparse, cuSPARSE and ViennaCL ({\tt 1.7.1}) on P100
\gpu{}s. Figure~\ref{fig:p100ppnr} shows the performance profile on P100 GPUs for NoReuse. 
Among these methods, {\sc kkspgemm} and cuSPARSE run for all $81$ instances. {\sc kkmem}, 
Nsparse and viennaCL fail for $2$, $4$ and $9$ matrices. cuSPARSE and
viennaCL are mostly outperformed by the other methods. 
%viennaCL and cuSPARSE achieve the best performance on $6$ and $1$ instances, respectively. 
These are followed by our previous method {\sc kkmem}, and our LP based method {\sc kklp}.
{\sc kkspgemm} takes advantage of {\sc kklp}, and significantly improves our previous method 
{\sc kkmem} with a better parameter setting. As a result, {\sc kkspgemm} and Nsparse are 
the most competitive methods. Nsparse, taking advantage of 
cuda-streams, achieves slightly better performance than {\sc kkspgemm}. Although the lack of 
cuda-streams is a limitation for {\sc kkspgemm}, with a better selection of the parameters it 
obtains the best performance for $28$ test problems. %(or at most $0.5\%$ slower than best KK method). 

\begin{figure}
\begin{center}
\includegraphics[width=0.60\columnwidth]{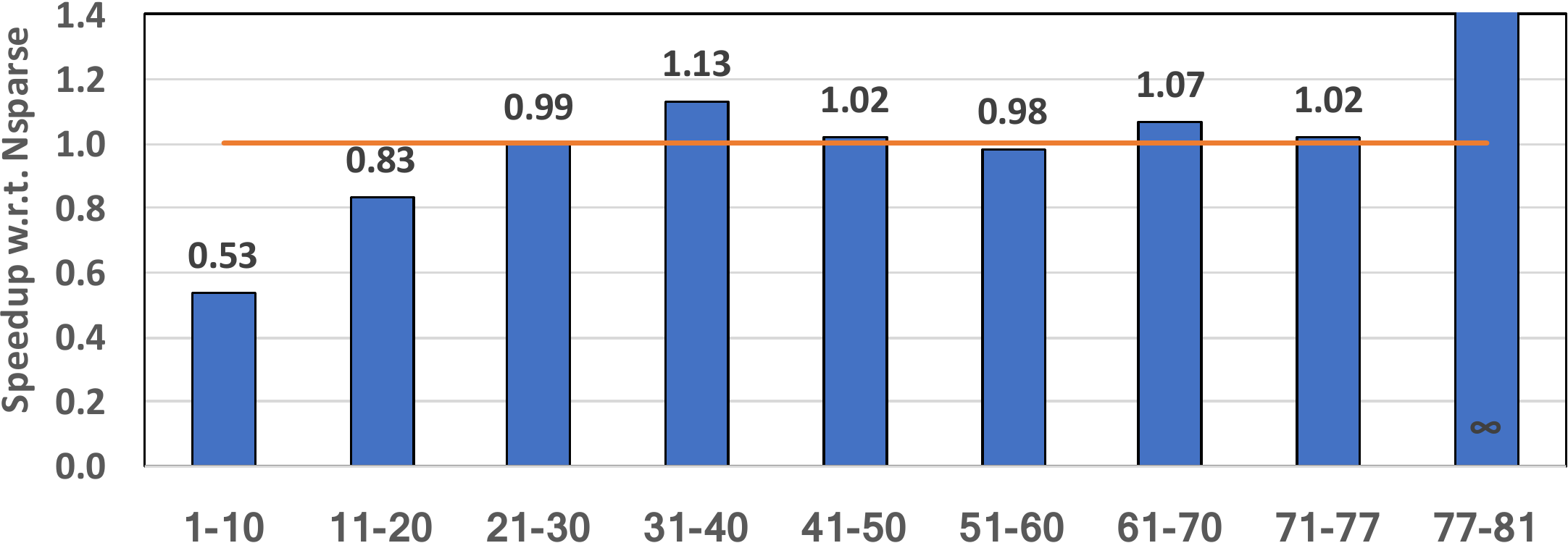}
\end{center}
\myspace{-3ex}
\caption{ \verysmallfont
Speedup of {\sc kkspgemm} w.r.t. NSparse for matrices that are grouped w.r.t. \flops{}.
These groups can be found using indices in Table~\ref{tab:overall1}. % and~\ref{tab:overall2}.
}
\label{fig:quantialespeedup}
\end{figure}

Most of the significant performance differences between Nsparse and {\sc kkspgemm}
occur for smaller multiplications that take between $1$ to $10$ milliseconds.
Nsparse has the best performance on $18$ out of $20$ multiplications 
with the smallest number of total \flops{}.
As the multiplications get larger, the performance of {\sc kkspgemm} is on average 
$3-4\%$ better than Nsparse (excluding the smallest $20$ test problems).
{\sc kkspgemm} is also able to perform $4$ test multiplications for which Nsparse runs out of memory %These include
(kron16, coPaparciteseer, flickr, coPapersDBLP). The performance comparison of {\sc kkspgemm}
against Nsparse for multiplications sorted based on \flops{} required 
is shown in Figure~\ref{fig:quantialespeedup}.
This figure reports the geometric mean of the {\sc kkspgemm} speedups w.r.t. 
Nsparse. For the smallest $10$ and $20$ multiplications, Nsparse is about $47\%$ and 
$17\%$ faster than {\sc kkspgemm}. {\sc kkspgemm}, on average, has more consistent and faster 
runtimes for the larger inputs. {\sc kkspgemm} is designed for scalability, and it introduces 
various overheads to achieve this scalability (e.g., compression). When the inputs are small, 
the overhead introduced is not amortized, as the multiplication time is very small 
even without compression. This makes {\sc kkspgemm} slower on small matrices, but at the same 
time it makes {\sc kkspgemm} more robust and scalable allowing it to run much larger problems.
On the other hand, Nsparse returns sorted output rows, which is not the case for {\sc kkspgemm}.
The choice of the better method depends on the application area. If the application
requires sorted outputs or the problem size is small, Nsparse is likely to achieve better performance.
For the problems with large memory requirements, {\sc kkspgemm} is the better choice.
Lastly, Figure~\ref{fig:p100ppr} gives the performance profile for the Reuse case. 
Although Nsparse also runs in two-phases, its current user interface does 
not allow reuse of the symbolic computations. 

\noindent
{\textbf {The effect of the compression:}} %The compression technique is applied in the {\tt symbolic phase} of {\sc kkmem}. 
Compression is critical to reduce the time and the memory requirements of the symbolic phase.
%as the row sizes are unknown, and they are estimated using \maxrowflops{}.
%This might cause memory problems on massively threaded architectures.
It helps to reduce both the number of hash insertions as well 
as the estimated max row size. 
Table-1 and 2 (supplementary materials) %~\ref{tab:overall1} and~\ref{tab:overall2} 
lists the original
\flops{} and \maxrowflops{}. CF and CMRF give the reduction ratios
with compression on \flops{} and \maxrowflops{} (e.g., $0.85$ means $15\%$ reduction), respectively. 
On average, \flops{} and \maxrowflops{} are reduced
by $62\%$ and $70\%$. 
Compression reduces the memory requirements (\maxrowflops{}) in most cases
up to $97\%$.
It usually reduces the runtime of the symbolic phase. 
When the reduction on \flops{} is low (e.g., CF $> 0.85$), it might not amortize compression cost.
We skip the second-phase of the compression in such cases; however, we still introduce 
overheads for CF calculations.
CF is greater than $0.85$ for only $7$ multiplications, for which the symbolic phase is 
run without compressed values.

\myspace{-3.5ex}
\section{Conclusion}
\myspace{-2ex}

We described thread-scalable \spgemm{} kernels for highly threaded architectures.
%We maintain performance portability in the kernel
%by handling underlying architecture limitations within a set of thread scalable
%data structures.
Using the portability provided by Kokkos, we describe algorithms
that are portable to \gpu{}s and \cpu{}s. The performance 
of the methods is demonstrated %against $1$ native method on Power8 \cpu{}s,
%$2$ on \knl{}s, and $3$ on P100 \gpu{}s, in which our implementations
on Power8 \cpu{}s, \knl{}s, and P100 \gpu{}s, in which our implementations
achieve at least as good performance as the native methods.
On \cpu{}s and \knl{}s, we show that sparse accumulators are preferrable 
when memory accesses are the performance bottleneck. As memory systems
provide more bandwidth (as in {\sc mcdram}) and $k$ is small, methods with dense 
accumulators outperform those with sparse accumulators.
Although our methods cannot exploit some of the architecture specific details of 
\gpu{}s, e.g., cuda-streams, because of current Kokkos limitations, with a 
better way of parameter selection we achieve as good performance as 
highly optimized libraries. The experiments also show that our methods 
using memory pool and compression techniques are robust and 
can perform multiplications with high memory demands. 
Our experiments also highlight the importance of designing methods 
for application use cases such as symbolic ``reuse'' with
significantly better performance than past methods.

{\verysmallerfont
\noindent {\bf Acknowledgements:} 
We thank %Grey Ballard, Erik Boman, Karen Devine, Carter Edwards, Simon Hammond,
Karen Devine % and Jonathan Berry % and Cynthia Phillips
for helpful discussions, and the test bed program at Sandia National Laboratories for
supplying the hardware used in this paper. 
Sandia National Laboratories is a multimission laboratory managed and
operated by National Technology and Engineering Solutions of Sandia,
LLC., a wholly owned subsidiary of Honeywell International, Inc., for
the U.S. Department of Energy's National Nuclear Security Administration
under contract DE-NA-0003525.}s

%This work is supported by the U.S. Dept. of Energy, 
%Office of Science, Office of Advanced Scientific Computing Research, Scientific 
%Discovery through Advanced Computing (SciDAC) program, and by the NNSA's 
%Advanced Simulation and Computing (ASC) program.


\begin{thebibliography}{10}

\bibitem{heroux2005overview}
The {T}rilinos project.

\bibitem{akbudak2014simultaneous}
Kadir Akbudak and Cevdet Aykanat.
\newblock Simultaneous input and output matrix partitioning for
  outer-product--parallel sparse matrix-matrix multiplication.
\newblock {\em SIAM Journal on Scientific Computing}, 36(5):C568--C590, 2014.

\bibitem{akbudak2017exploiting}
Kadir Akbudak and Cevdet Aykanat.
\newblock Exploiting locality in sparse matrix-matrix multiplication on
  many-core architectures.
\newblock {\em IEEE Transactions on Parallel and Distributed Systems}, 2017.

\bibitem{azad2015exploiting}
Ariful Azad, Grey Ballard, Aydin Buluc, James Demmel, Laura Grigori, Oded
  Schwartz, Sivan Toledo, and Samuel Williams.
\newblock Exploiting multiple levels of parallelism in sparse matrix-matrix
  multiplication.
\newblock {\em arXiv preprint arXiv:1510.00844}.

\bibitem{ballard2016hypergraph}
Grey Ballard, Alex Druinsky, Nicholas Knight, and Oded Schwartz.
\newblock Hypergraph partitioning for sparse matrix-matrix multiplication.
\newblock {\em arXiv preprint arXiv:1603.05627}, 2016.

\bibitem{bulucc2011combinatorial}
Ayd{\i}n Bulu{\c{c}} and John~R Gilbert.
\newblock {The Combinatorial BLAS: Design, implementation, and applications}.
\newblock {\em International Journal of High Performance Computing
  Applications}, 2011.

\bibitem{cohen1998structure}
Edith Cohen.
\newblock Structure prediction and computation of sparse matrix products.
\newblock {\em Journal of Combinatorial Optimization}, 2(4):307--332, 1998.

\bibitem{dalton2015optimizing}
Steven Dalton, Luke Olson, and Nathan Bell.
\newblock Optimizing sparse matrix-—matrix multiplication for the {GPU}.
\newblock {\em ACM Transactions on Mathematical Software (TOMS)}, 41(4):25,
  2015.

\bibitem{davis2011university}
Timothy~A Davis and Yifan Hu.
\newblock {The University of Florida sparse matrix collection}.
\newblock {\em ACM Transactions on Mathematical Software (TOMS)}, 38(1):1,
  2011.

\bibitem{demouth2012sparse}
Julien Demouth.
\newblock Sparse matrix-matrix multiplication on the gpu.
\newblock In {\em Proceedings of the GPU Technology Conference}, 2012.

\bibitem{deveci16coloring}
M.~Deveci, E.~G. Boman, K.~D. Devine, and S.~Rajamanickam.
\newblock Parallel graph coloring for manycore architectures.
\newblock In {\em 2016 IEEE International Parallel and Distributed Processing
  Symposium (IPDPS)}, pages 892--901, May 2016.

\bibitem{deveci2013hypergraph}
Mehmet Deveci, Kamer Kaya, and Umit~V Catalyurek.
\newblock Hypergraph sparsification and its application to partitioning.
\newblock In {\em Parallel Processing (ICPP), 2013 42nd International
  Conference on}, pages 200--209. IEEE, 2013.

\bibitem{deveci2017performance}
Mehmet Deveci, Christian Trott, and Sivasankaran Rajamanickam.
\newblock Performance-portable sparse matrix-matrix multiplication for
  many-core architectures.
\newblock In {\em Parallel and Distributed Processing Symposium Workshops
  (IPDPSW), 2017 IEEE International}, pages 693--702. IEEE, 2017.

\bibitem{dysart2016highly}
Timothy Dysart, Peter Kogge, Martin Deneroff, Eric Bovell, Preston Briggs, Jay
  Brockman, Kenneth Jacobsen, Yujen Juan, Shannon Kuntz, Richard Lethin, et~al.
\newblock Highly scalable near memory processing with migrating threads on the
  emu system architecture.
\newblock In {\em Irregular Applications: Architecture and Algorithms (IA3),
  Workshop on}, pages 2--9. IEEE, 2016.

\bibitem{edwards2014kokkos}
H~Carter Edwards, Christian~R Trott, and Daniel Sunderland.
\newblock Kokkos: Enabling manycore performance portability through polymorphic
  memory access patterns.
\newblock {\em J Parallel Distrib Comp}, 74(12):3202--3216, 2014.

\bibitem{gremse2015gpu}
Felix Gremse, Andreas Hofter, Lars~Ole Schwen, Fabian Kiessling, and Uwe
  Naumann.
\newblock {GPU}-accelerated sparse matrix-matrix multiplication by iterative
  row merging.
\newblock {\em SIAM Journal on Scientific Computing}, 37(1):C54--C71, 2015.

\bibitem{gustavson1978two}
Fred~G Gustavson.
\newblock Two fast algorithms for sparse matrices: Multiplication and permuted
  transposition.
\newblock {\em ACM Transactions on Mathematical Software (TOMS)},
  4(3):250--269, 1978.

\bibitem{intel2007intel}
Intel.
\newblock Intel math kernel library, 2007.

\bibitem{kunchum2017improving}
Rakshith Kunchum, Ankur Chaudhry, Aravind Sukumaran-Rajam, Qingpeng Niu, Israt
  Nisa, and P~Sadayappan.
\newblock On improving performance of sparse matrix-matrix multiplication on
  gpus.
\newblock In {\em Proceedings of the International Conference on
  Supercomputing}, page~14. ACM, 2017.

\bibitem{kurt2017Characterization}
Sureyya~Emre Kurt, Vineeth Thumma, Changwan Hong, Aravind Sukumaran-Rajam, and
  P~Sadayappan.
\newblock Characterization of data movement requirements for sparse matrix
  computations on gpus.
\newblock In {\em High Performance Computing (HiPC), 2017 24th International
  Conference on}. IEEE, 2017.

\bibitem{lin2014towards}
Paul Lin, Matthew Bettencourt, Stefan Domino, Travis Fisher, Mark Hoemmen,
  Jonathan Hu, Eric Phipps, Andrey Prokopenko, Sivasankaran Rajamanickam,
  Christopher Siefert, et~al.
\newblock Towards extreme-scale simulations for low mach fluids with
  second-generation trilinos.
\newblock {\em Parallel Processing Letters}, 24(04):1442005, 2014.

\bibitem{liu2014efficient}
Weifeng Liu and Brian Vinter.
\newblock An efficient gpu general sparse matrix-matrix multiplication for
  irregular data.
\newblock In {\em Parallel and Distributed Processing Symposium, 2014 IEEE 28th
  International}, pages 370--381. IEEE, 2014.

\bibitem{mccourt2013efficient}
Michael McCourt, Barry Smith, and Hong Zhang.
\newblock Efficient sparse matrix-matrix products using colorings.
\newblock {\em SIAM Journal on Matrix Analysis and Applications}, 2013.

\bibitem{nagasaka2017high}
Yusuke Nagasaka, Akira Nukada, and Satoshi Matsuoka.
\newblock High-performance and memory-saving sparse general matrix-matrix
  multiplication for nvidia pascal gpu.
\newblock In {\em Parallel Processing (ICPP), 2017 46th International
  Conference on}, pages 101--110. IEEE, 2017.

\bibitem{naumov2010cusparse}
M~Naumov, LS~Chien, P~Vandermersch, and U~Kapasi.
\newblock Cusparse library.
\newblock In {\em GPU Technology Conference}, 2010.

\bibitem{patwary2015parallel}
Md~Mostofa~Ali Patwary, Nadathur~Rajagopalan Satish, Narayanan Sundaram,
  Jongsoo Park, Michael~J Anderson, Satya~Gautam Vadlamudi, Dipankar Das,
  Sergey~G Pudov, Vadim~O Pirogov, and Pradeep Dubey.
\newblock Parallel efficient sparse matrix-matrix multiplication on multicore
  platforms.
\newblock In {\em High Performance Computing}, pages 48--57. Springer, 2015.

\bibitem{Rupp:ViennaCL}
K.~Rupp, F.~Rudolf, and J.~Weinbub.
\newblock {ViennaCL - A High Level Linear Algebra Library for GPUs and
  Multi-Core CPUs}.
\newblock In {\em Intl.~Workshop on GPUs and Scientific Applications}, pages
  51--56, 2010.

\bibitem{wolf2017fast}
Michael~M Wolf, Mehmet Deveci, Jonathan~W Berry, Simon~D Hammond, and
  Sivasankaran Rajamanickam.
\newblock Fast linear algebra-based triangle counting with kokkoskernels.
\newblock In {\em High Performance Extreme Computing Conference (HPEC), 2017
  IEEE}, pages 1--7. IEEE, 2017.

\end{thebibliography}
\end{document}